\documentclass[10pt,journal,compsoc]{IEEEtran}
\usepackage{amsmath,amsfonts}
\usepackage{algorithmic}
\usepackage{array}
\usepackage[caption=false,font=normalsize,labelfont=sf,textfont=sf]{subfig}
\usepackage{textcomp}
\usepackage{stfloats}
\usepackage{url}
\usepackage{verbatim}
\usepackage{graphicx}
\usepackage{cite}

\usepackage{enumitem}
\usepackage{colortbl}
\usepackage{makecell}
\usepackage{wrapfig}
\usepackage{tabularx,booktabs}

\usepackage{ifpdf}
 \ifpdf
 \else
 \fi

\makeatletter
\let\NAT@parse\undefined
\makeatother
\usepackage{hyperref}  
\def\BibTeX{{\rm B\kern-.05em{\sc i\kern-.025em b}\kern-.08em
    T\kern-.1667em\lower.7ex\hbox{E}\kern-.125emX}}
\usepackage{balance}

\begin{document}
\title{A Survey of Developable Surfaces: From Shape Modeling to Manufacturing}

\author{Chao~Yuan,~Nan~Cao,~and~Yang~Shi
\IEEEcompsocitemizethanks{
\IEEEcompsocthanksitem Chao Yuan, Nan Cao, and Yang Shi are with the Intelligent Big Data Visualization Lab, Tongji University, Shanghai, China. Nan Cao and Yang Shi are the corresponding author. E-mail: nan.cao@gmail.com, and yangshi.idvx@tongji.edu.cn}}


\IEEEtitleabstractindextext{
\begin{abstract}
Developable surfaces are commonly observed in various applications such as architecture, product design, manufacturing, mechanical materials, and data physicalization as well as in the development of tangible interaction and deformable robots, with the characteristics of easy-to-product, low-cost, transport-friendly, and deformable. Transforming shapes into developable surfaces is a complex and comprehensive task, which forms a variety of methods of segmentation, unfolding, and manufacturing for shapes with different geometry and topology, resulting in the complexity of developable surfaces. In this paper, we reviewed relevant methods and techniques for the study of developable surfaces, characterize them with our proposed pipeline, and categorize them based on digital modeling, physical modeling, interaction, and application. Through the analysis to the relevant literature, we also discussed some of the research challenges and future research opportunities.
\end{abstract}

\begin{IEEEkeywords}
developable surfaces, developable approximation, digital fabrication, physicalization.
\end{IEEEkeywords}
}
\maketitle
\IEEEdisplaynontitleabstractindextext
\IEEEpeerreviewmaketitle

\section{Introduction}

\IEEEPARstart{D}EVELOPABLE surfaces are widely used in construction, complex shape manufacturing, garment manufacturing, and deformable mechanical materials, where they have advantages over traditional manufacturing methods \cite{luo2023autonomous}. Through processing planar materials (bending, folding, rolling, cutting, assembling, etc.), complex 3D shapes can be fabricated, which provides enhanced design flexibility. The idea of forming 3D shapes from 2D planar materials has many advantages, such as reducing manufacturing difficulty, saving materials, and lowering transportation costs. In addition, developable surfaces are closely related to flexible manufacturing \cite{QIN2016173}, which can improve the efficiency and intelligence of manufacturing.

Developable surfaces can be formally defined as surfaces with zero Gaussian curvature. According to the definition of Gaussian curvature, at least one principal curvature is zero, so all developable surfaces in 3D space are ruled surfaces, such as cylindrical surfaces, conical surfaces, tangent surfaces, and planes. Developable surfaces can be flattened without distortion into a plane, and the area of the unfolded plane is the same as that of the original surface; only their coordinates are transformed. To study developable surfaces, planes can be formed into surfaces and complex shapes through folding, such as in origami. This design approach has inspired many structural innovations, such as folding architecture and solar panels on satellites. Additionally, transforming 3D shapes into developable surfaces is a general forming method that can be used in more fields. This study focuses on the transformation from 3D shapes to developable surfaces.

However, the process of transforming 3D shapes into developable surfaces is a complex and multifaceted task, presenting numerous challenges at various stages. The diversity of design requirements contributes to the wide range of geometry and topology found in target shapes, thereby increasing the difficulty of the operation. Different segmentation layouts resulting from the various segmentation methods employed can significantly impact subsequent flattening processes and manufacturing complexities. When it comes to shape unfolding computation, deformation errors can arise due to factors such as the topological structure of the shape, computation time, computational power, and other limitations. Furthermore, the results of these transformations are subject to manufacturing and assembly process restrictions. These examples illustrate just a few of the challenges inherent in the transformation of 3D shapes into developable surfaces.

\begin{figure*}[!t]
    \centering
    \includegraphics[width=\textwidth]{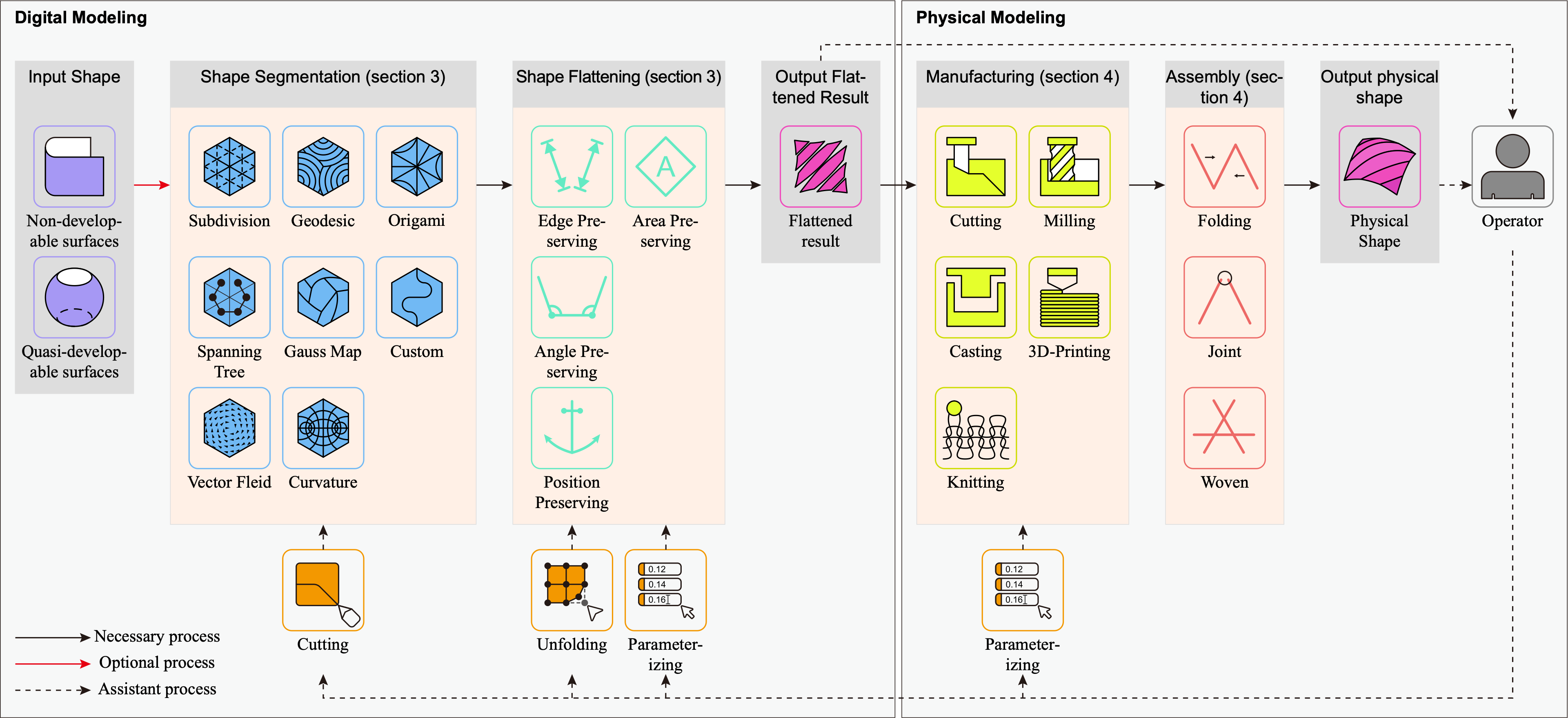}
    \vspace{-20pt}
    \caption{Pipeline of developable surfaces from digital modeling to physical modeling. Shape segmentation: divide input 3D shapes into several or many pieces based on certain rules. Surface flattening: flatten a surface into a plane with minimum error. Output flattened result: digital flat surfaces. Manufacturing: produce physical developable surfaces based on the digital model. Assembly: Assembly of developable surfaces. Interactive assistance: alternative methods for better or more convenient optimization of the deformation process.}
    \label{pipeline}
\end{figure*}

These difficulties and diverse methodological challenges have motivated extensive investigation aiming to address one or more aspects of developable surface. Some studies propose more general unfolding methods that can be applied to different target shapes. Other studies focus on specific types of tasks and provide one or more unfolding solutions. They have led to research based on specific fields, such as architecture and manufacturing, where the characteristics of developable surfaces can simplify manufacturing difficulties and costs. In the study of new mechanical materials, the deformability of developable surfaces contributes to the exploration of programmable mechanical materials. These studies provide methods that not only effectively simplify the manufacturing difficulties of complex shapes but also lead to a surge in new design directions and generated styles.
The purpose of this survey is to provide a comprehensive overview of methods for segmentation and flattening and research on applications based on developable surfaces. By collecting and analyzing relevant literature on the topic, we identify the relevant processes and operations of developable surfaces and summarize a pipeline (see Fig. \ref{pipeline}) and analyze the various techniques used in the relevant processes. Based on different types of target shapes, we use the pipeline to organize the corresponding processes and operations and determine a general unfolding method for target shapes. In addition, we explored research directions and applications that have received less attention in previous work and remain challenges for future research.

\section{Related Surveys and Methodology}

In this section, we first review surveys relevant to fields of developable surfaces. After that, we introduce our method for collecting and categorizing papers.

\subsection{Related Surveys}

This section collects the recent surveys on developable surfaces. These surveys are divided into two categories. One focuses on general methods and techniques of developable surfaces, and the other focuses on applications based on folding forms.

In the first type, Bhanage reviewed developable surfaces in the field of sheet metal \cite{bhanage_overview_2014}. Nejur and Steinfeld reviewed some mesh segmentation algorithms by mesh dual graph applied in “generative architectural design” \cite{nejur_bringing_2016}. Stein analyzed "several smoothness energies in geometry processing" and presented new discrete developability methods \cite{stein_smoothness_2020}. Zhang and Zheng reviewed and classified developable surface techniques \cite{zhang_overview_2022}. Compared with the above review of developable surfaces, our survey provides a more holistic overview of analytics approaches for the process from discrete developability to physical modeling, the methods and ideas used, and the related application fields.

In the second type, origami structure, a special developable surface, can be used in a variety of fields. In mechanical engineering, origami structures offer unique material properties. Peraza-Hernandez et al. reviewed “active materials” based on origami \cite{peraza-hernandez_origami-inspired_2014}. Turner et al. reviewed mechanical devices based on origami applications \cite{turner_review_2016}. Johnson et al. proposed an overview of applications in biomedical devices by origami structures \cite{johnson_fabricating_2017}. Li et al. focused on the research of geometry and properties of materials by origami \cite{li_architected_2019}. Shah et al. reviewed deployable antennas based on origami methods \cite{shah_lightweight_2021}. Meloni et al. presented origami methods for engineering applications from 2015 to 2020 \cite{meloni_engineering_2021}. Fonseca et al. proposed an overview of “origami-inspired systems and structures” for smart materials \cite{fonseca_overview_2022}. Some reviews focus on architectural applications. Doroftei et al. focused on the overview of applications of “foldable plate structures” in architecture \cite{doroftei_overview_2018}. In product design, Meloni et al. reviewed product designs based on origami structures \cite{meloni_engineering_2021}. In the field of origami art and design, Demaine et al. reviewed curved folding in art and design \cite{demaine_review_2015}. Based on these prior works, we aim to provide a comprehensive overview of research based on developable surfaces.

\subsection{Survey Methodology and Taxonomy}

To provide a comprehensive review of existing studies, we collected relevant papers from computer graphics journals and conferences by using two main approaches: search-driven and citation-driven selection. For the former, we first used a keyword approach in the ACM literature search database to obtain the initial papers based on the extended schema of the ACM Guide to Computing Literature. The relevant keywords have multiple expressions, so we listed as many expressions as possible and connected these words by logical "OR" to obtain a total of 622 initial papers. We then reviewed the abstracts of each paper and then filtered out the papers that did not meet the requirements. The reason is that these papers only mention the keywords of "developable surfaces" but not the study of developable surfaces. For the citation-driven selection, we extended the collection by using publications that we knew in advance about the topic and relevant citations from the articles collected based on the search drive. In addition to the above search, some relevant studies were included. We reviewed papers on relevant AI techniques for developable surfaces, but studies on developable surfaces based on AI techniques are few. For shape segmentation methods, we also searched for corresponding papers based on keywords, which involve many deep learning techniques but target semantic-based shape segmentation. Finally, we added studies on the interaction of developable surfaces by keyword search, which obtained relevant examples of tangible interaction studies. The following list shows these keywords and their alternative expressions:
\begin{itemize}[leftmargin=*]
\item {\textbf{Developable surfaces}}: "developable surface", "developable surfaces", "developable structure", "developable structures", origami and kirigami.
\item {\textbf{AI}}: "AI", "artificial intelligence", "machine learning", "deep learning", "reinforcement learning", "convolutional neural network", "CNN", "CNNs", "style transfer", "Generative adversarial network", "GAN" OR "variational autoencoder", "VAE", "autoencoder", "ANN", "graph neural network", "GNN".
\item {\textbf{Mesh segmentation}}: "mesh segmentation", "mesh separation", "meshes segmentation", "meshes separation".
\end{itemize}

We eventually collected a total of 135 articles (2010–2022), including 121 papers on developable surfaces and 14 papers on mesh segmentation. These papers concentrate on \textit{ACM TOG} (54 papers), \textit{Computer-Aided Design} (14 papers), \textit{Computers \& Graphics} (7 papers), \textit{CHI} (6 papers), \textit{SimAUD} (3 papers), and \textit{Computer Graphics Forum} (3 papers).

\begin{figure*}[t!]
    \centering
    \includegraphics[width=\textwidth]{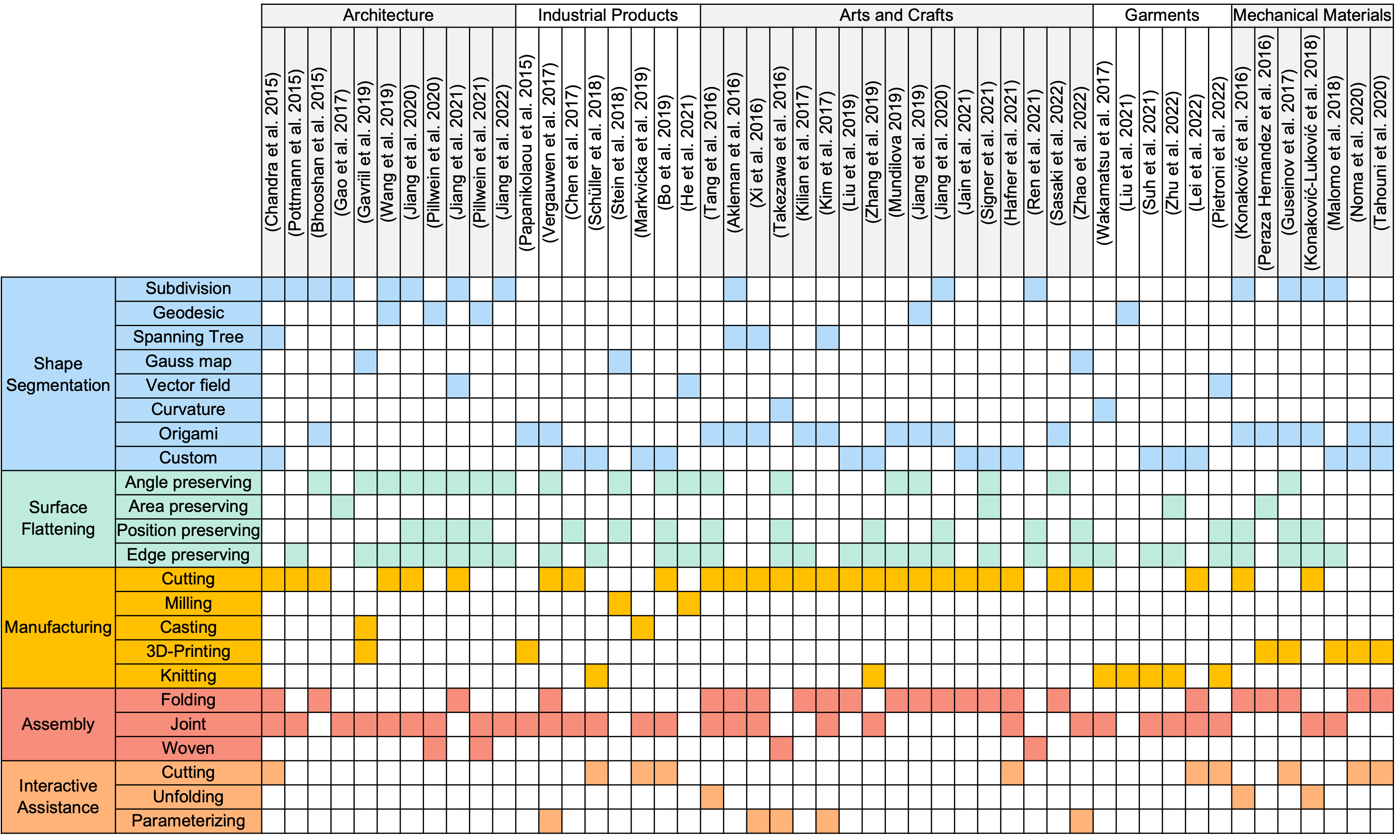}
    \vspace{-20pt}
    \caption{Visual Analysis of Highly Cited Studies between 2015 and 2022 according to collected research.}
    \label{matrix}
\end{figure*}

We summarized a comprehensive pipeline of developable surfaces based on the research process of analyzing the collected literature. The pipeline is divided into two stages: digital modeling and physical modeling. Owing to the geometric and topological diversity of the shape,  developable approximation methods are different. The purpose of the digital modeling stage is to transform 3D shapes into unfolded planes. We subdivided the digital modeling stage into two key processes: shape segmentation and surface flattening. Some interactive assistance that connects the digital modeling stage is also necessary, as the operator better adjusts the details of the shapes during the segmentation and flattening based on the feedback of the output visual results. 
On the contrary, the purpose of the physical modeling stage is to produce the corresponding physical shape from the data of the digital model and output the final result through the corresponding assembly. We subdivided the physical forming stage into two key processes: manufacturing and assembly. According to the physical modeling result, the operator can also adjust parameters of numerical control equipment reducing errors. Each process of the pipeline is independent, and these processes as different dimensions form a design space (see Fig. \ref{matrix}), where each technique reflects one or more design choices in each dimension.

The rest of the contents are organized as follows. We analyze the corresponding techniques in the digital modeling stage in section\ref{sec3}. The techniques for shape segmentation are analyzed and summarized in Section\ref{sec3.1}, and the techniques for surface flattening are analyzed and summarized in Section\ref{sec3.2}. Section\ref{sec4} analyzes the physical modeling methods, including manufacturing techniques and assembly methods. Then, Section\ref{sec5} summarizes the interactive assistance in the shape segmentation and flattening. We also analyze the association between digitization and physicalization and the corresponding interaction. Section\ref{sec6} summarizes some of the application areas involved in developable surfaces. Finally, we discuss challenges and opportunities in the study of developable surfaces in Section\ref{sec7} and conclude in Section\ref{sec8}.

\section{Digital Modeling}\label{sec3}

The purpose of digital modeling is to transform 3D shapes into unfolded planes. Given the non-zero Gaussian curvature of each point on the surface of most 3D shapes, most 3D shapes are not directly unfolded. Thus, a 3D shape is usually segmented into many parts and then flattened into flat surfaces. We divide digital modeling into two types of operations: \textbf{shape segmentation} for segmentation of 3D shapes and \textbf{surface flattening} for unfolding surfaces.

\subsection{Shape Segmentation}\label{sec3.1}

In this section, we introduce segmentation operations of undevelopable shapes. Shape segmentation is dividing a shape into pieces that are easy to unfold. How to set the segmentation operation reasonably is particularly important for better flattening the surface and reducing the error between the target shape and generated result. We can divide shape segmentation into seven methods (table \ref{seg}): \textbf{subdivision} for fitting target shapes by discrete mesh, \textbf{geodesic} for segmentation by shortest path on a surface, \textbf{spanning tree} for segmentation by dual graph, \textbf{Gauss map} for generation of creases by dealing with the normals of Gaussian image, \textbf{vector field} for reconstruction of the grid, \textbf{curvature} for reconstruction and segmentation of mesh, \textbf{origami} for generation of creases based on mesh edges; and \textbf{custom} for user-defined operation. During shape segmentation, some results are formed using a single method, whereas others are formed by multiple methods.

\begin{table*}[t!]
    \centering
    \caption{Comparison of different shape segmentation.}
    \label{seg}
    \vspace{-5pt}
    {
        \begin{tabular}{|ccccr|}
            \hline
            Segmentation & Characteristics & Effect & Available Techniques & Examples \\ \hline
            
            \makecell[c]{\begin{minipage}[b]{0.12\columnwidth}
            \raisebox{-0.8\height}{\includegraphics[width=\linewidth]{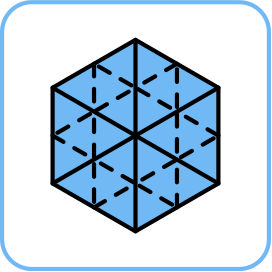}}
            \end{minipage} \\ Subdivision} & 
            \begin{minipage}{0.2\textwidth}
                {Polygonal dividing lines formed based on mesh subdivision}
            \end{minipage} & 
            \begin{minipage}{0.2\textwidth}
                {Subdivide shapes into a large number of parts which approximate large-scale shapes}
            \end{minipage} & 
            \makecell[c]{Loop subdivision\\Catmull–Clark \\Parametric domain\\Isometric mapping\\Circle parking\\Mesh dual\\Clustering} & 
            \makecell[r]{\cite{erdine_curved-crease_2019} \\ \cite{tang_form-finding_2014} \\ \cite{gale_patterning_2016} \\ \cite{jiang_using_2021} \\ \cite{pottmann_cell_2015} \\ \cite{pellis_principal_2020} \\ \cite{fu_k-set_2010}} \\ \hline
            
            \makecell[c]{\begin{minipage}[b]{0.12\columnwidth}
            \raisebox{-0.8\height}{\includegraphics[width=\linewidth]{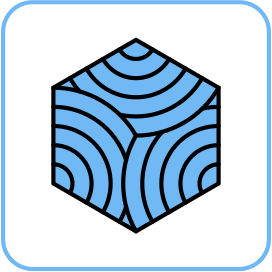}}
            \end{minipage} \\ Geodesic} & 
            \begin{minipage}{0.2\textwidth}
                {Partially parallel-like dividing lines with intersection formed based on geodesic}
            \end{minipage} & 
            \begin{minipage}{0.2\textwidth}
                {Improve the continuity of the surface by bending the material, which reduce the number of building units}
            \end{minipage} & 
            \makecell[c]{Constant width geodesic\\Distance field} & \makecell[r]{\cite{pottmann_geodesic_2010} \\ \cite{pillwein_elastic_2020}} \\ \hline
            
            \makecell[c]{\begin{minipage}[b]{0.12\columnwidth}
            \raisebox{-0.8\height}{\includegraphics[width=\linewidth]{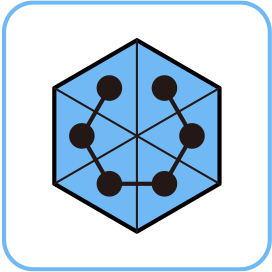}}
            \end{minipage} \\ Spanning tree} & 
            \begin{minipage}{0.2\textwidth}
                {Dividing lines formed by the nodes that are not connected in the dual graph of the mesh}
            \end{minipage} & 
            \begin{minipage}{0.2\textwidth}
                {Provide a method of constructing polygonal shapes by strip folding}
            \end{minipage} & \makecell[c]{Mesh dual graph \\ Greedy algorithms \\ Clustering \\ Genetic algorithm} & \makecell[r]{\cite{leung_prototyping_2018} \\ \cite{chandra_computing_2015} \\ \cite{xi_learning_2016} \\ \cite{kim_disjoint_2017}} \\ \hline
            
            \makecell[c]{\begin{minipage}[b]{0.12\columnwidth}
            \raisebox{-0.8\height}{\includegraphics[width=\linewidth]{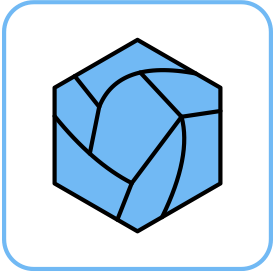}}
            \end{minipage} \\ Gauss map} & 
            \begin{minipage}{0.2\textwidth}
                {Irregular dividing lines formed based on Gauss map}
            \end{minipage} & 
            \begin{minipage}{0.2\textwidth}
                {Divide and optimize complex shapes into developable surfaces}
            \end{minipage} &
            \makecell[c]{Clustering \\ Minimum energy} & \makecell[r]{\cite{tang_optimization_2005} \\ \cite{zhao_developability-driven_2022}} \\ \hline
            
            \makecell[c]{\begin{minipage}[b]{0.12\columnwidth}
            \raisebox{-0.8\height}{\includegraphics[width=\linewidth]{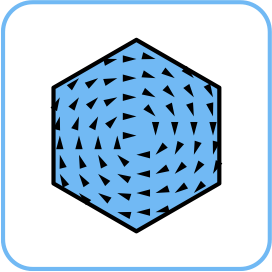}}
            \end{minipage} \\ Vector field} & 
            \begin{minipage}{0.2\textwidth}
                {Polygonal dividing lines formed by guidance of vector field}
            \end{minipage} & 
            \begin{minipage}{0.2\textwidth}
                {Guide generation of dividing lines and mesh}
            \end{minipage} & \makecell[c]{Minimum energy} & \makecell[r]{\cite{verhoeven_dev2pq_2022} \\ \cite{pietroni_computational_2022}} \\ \hline
            
            \makecell[c]{\begin{minipage}[b]{0.12\columnwidth}
            \raisebox{-0.8\height}{\includegraphics[width=\linewidth]{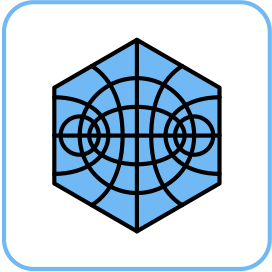}}
            \end{minipage} \\ Curvature} & 
            \begin{minipage}{0.2\textwidth}
                {Orthogonal dividing lines formed based on curvature}
            \end{minipage} & 
            \begin{minipage}{0.2\textwidth}
                {Automatically subdivide shapes into a large number of parts by the generated orthogonal network}
            \end{minipage} & \makecell[c]{principal curvature \\ Gaussian curvature} & \makecell[r]{\cite{takezawa_fabrication_2016} \\ \cite{chen_g2_2013}} \\ \hline
            
            \makecell[c]{\begin{minipage}[b]{0.12\columnwidth}
            \raisebox{-0.8\height}{\includegraphics[width=\linewidth]{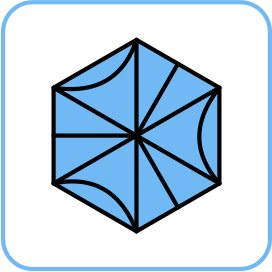}}
            \end{minipage} \\ Origami} & 
            \begin{minipage}{0.2\textwidth}
                {Dividing lines formed by creases from mesh edges}
            \end{minipage} & 
            \begin{minipage}{0.2\textwidth}
                {Provides methods to fast build 3D shapes by folding}
            \end{minipage} & \makecell[c]{Tuck-folding} & \makecell[r]{\cite{tachi_origamizing_2010} \\ \cite{jiang_freeform_2020}} \\ \hline
            
            \makecell[c]{\begin{minipage}[b]{0.12\columnwidth}
            \raisebox{-0.8\height}{\includegraphics[width=\linewidth]{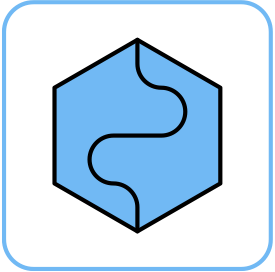}}
            \end{minipage} \\ Custom} & 
            \begin{minipage}{0.2\textwidth}
                {Dividing lines formed by user-defined segmentation}
            \end{minipage} & 
            \begin{minipage}{0.2\textwidth}
                {Provide the operator with a custom method of splitting shapes}
            \end{minipage} & \makecell[c]{manual operation} & \makecell[r]{\cite{skouras_designing_2014} \\ \cite{schuller_shape_2018}} \\ \hline
        \end{tabular}
    }
\end{table*}

\textbf{Subdivision.}
Subdivision is the division of a target shape into many similar and regular meshes, and then each mesh is discretized into a planar mesh. The method is often used in the field of large-scale construction whose each part is convenient to fabrication. According to the shape of the mesh, we can divide the results of subdivision into the triangular mesh, quad mesh, and polygon mesh. These mesh types have corresponding properties.

\begin{figure*}[t!]
    \centering
    \includegraphics[width=\textwidth]{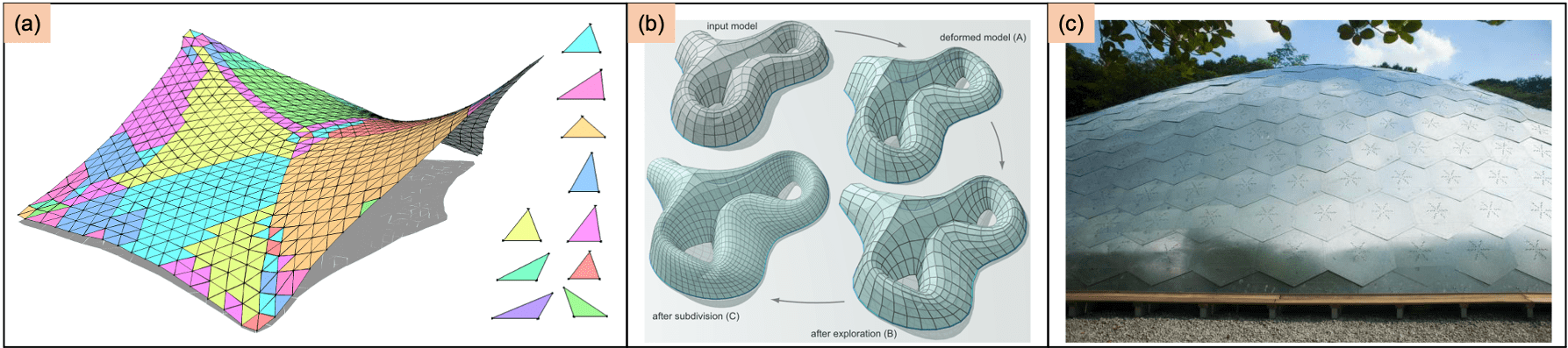}
    \vspace{-20pt}
    \caption{3D shapes segmentation by subdivision: (a) Triangular meshes clustering \cite{singh_triangle_2010}. (b) Planar quad mesh subdivision \cite{yang_shape_2011}. (c) SUTD pavilion construction by planar polygon meshes \cite{sevtsuk_freeform_2014}.}
    \label{subd}
\end{figure*}

Triangular mesh is a planar mesh according to its geometric properties, which is the most intuitive method to directly discretize undevelopable shapes into many triangular meshes \cite{erdine_curved-crease_2019}. The triangular mesh can be obtained directly by loop subdivision \cite{stam1998evaluation}, so much research focuses on how to reassemble easily. Some studies propose methods of connecting every triangular mesh by setting additional meshes. For instance, each mesh face of the original shape is connected by use of folding and tucking additional meshes, which makes each mesh face reassembled into a 3D shape by folding \cite{tachi_origamizing_2010, peraza_hernandez_smi_2013, guseinov_curveups_2017}. To analyze the similarity of each mesh face, K-means clustering is a suitable approach that reduces the variety of mesh face types and mitigates manufacturing complexities \cite{singh_triangle_2010} (see Fig. \ref{subd}a). For triangular mesh deformation, triangular meshes approximate a freeform shape with “curved-folded tessellation” \cite{chandra_computing_2015}. Triangular meshes with cutting and stretching can approximate a 3D shape by rotating every mesh face \cite{konakovic_beyond_2016, konakovic-lukovic_rapid_2018}.

\begin{figure}[ht]
    \centering
    \vspace{-10pt}
    \includegraphics[width=0.75\linewidth]{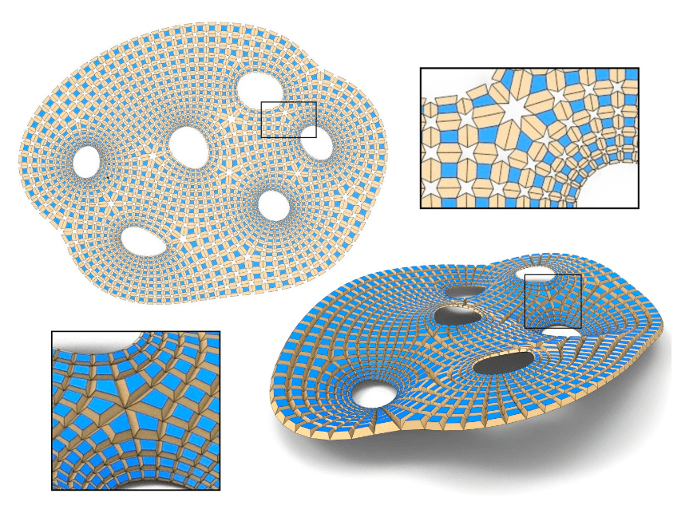}
    \vspace{-18pt}
    \caption{Creases generation by tuck-folding of origami \cite{jiang_freeform_2020}}
    \label{tuck}
\end{figure}

Quad mesh with a regular and beautiful structure is widely used in many fields. Quad mesh subdivision is a widely used method to generate quad mesh \cite{catmull_recursively_1978}, and a polygonal mesh is formed by subdivision \cite{tang_form-finding_2014, jiang_quad-mesh_2020}. Some studies are about folding polygonal shapes by origami \cite{tachi_origamizing_2010, jiang_freeform_2020} (see Fig. \ref{tuck}). Subdivided quad mesh is generated by UV nets of parametric surface \cite{gale_patterning_2016}, by mapping parametric domain \cite{gao_grid_2017}, and by isometric mapping \cite{jiang_quad-mesh_2020, jiang_using_2021}. K-means clustering is used to analyze the similarity of quad mesh  \cite{eigensatz_paneling_2010, fu_k-set_2010}. In addition, spatial quad mesh faces are optimized into planar quad mesh faces, which can approximate a 3D shape \cite{bo_circular_2011, yang_shape_2011, vouga_design_2012, jiang_freeform_2014, akleman_construction_2016, jiang_freeform_2020, jiang_planar_2022} (see Fig. \ref{subd}b).

Polygon mesh is over the four edges. Various studies focus on approximating 3D shapes using polygon meshes. For example, there are techniques for generating hexagonal meshes aligned with planar quad meshes \cite{jiang_freeform_2014}, polygon mesh generation through circle packing methods \cite{pottmann_cell_2015}, and construction of polygon meshes using mesh dual approaches \cite{bo_circular_2011, vouga_design_2012, jiang_polyhedral_2015, pellis_principal_2020}. Additionally, polygon meshes are employed in the construction of freeform formwork \cite{sevtsuk_freeform_2014, ren_3d_2021} (see Fig. \ref{subd}c).

\textbf{Geodesic.}
A geodesic line is the shortest path between two points on a local surface, which can be approximated locally as a straight line. A regular net can be generated by geodesic lines, and relevant studies improve the basic method for different 3D shapes.

\begin{figure*}[t!]
    \centering
    \includegraphics[width=\textwidth]{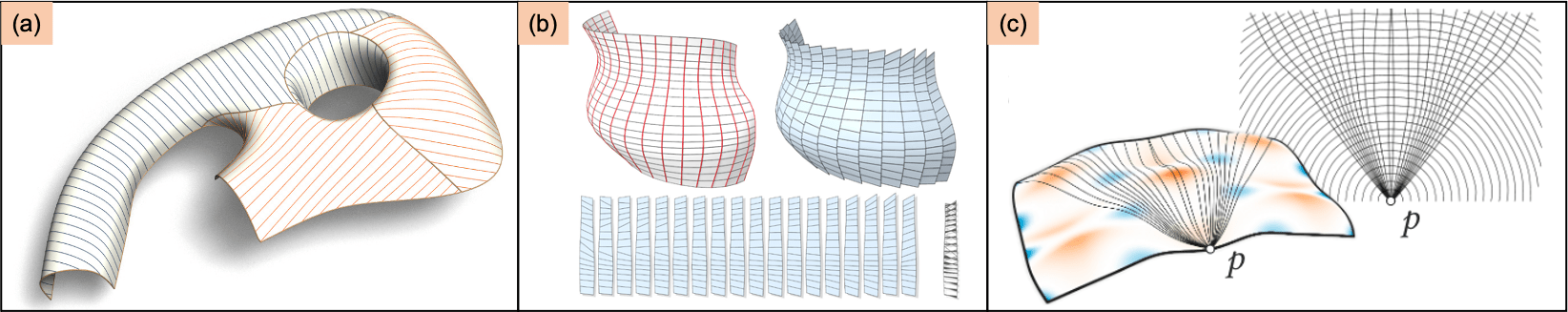}
    \vspace{-20pt}
    \caption{3D shapes segmentation based on geodesic: (a) Constant width geodesic curves on surfaces \cite{pottmann_geodesic_2010}. (b) Surfaces split by orthogonal nets \cite{wang_discrete_2019}. (c) Constructing shortest geodesic by distance fields \cite{pillwein_elastic_2020}.}
    \label{geodesic}
\end{figure*}

Constant width nets are formed by equidistant geodesic lines to divide the target shape into uniform strips. For instance, a divided net is formed by geodesic lines with constant width onto surface \cite{pottmann_geodesic_2010} (see Fig. \ref{geodesic}a) and by “pseudo-geodesics” (not strict geodesics lines) \cite{jiang_curve-pleated_2019}.

Orthogonal nets refer to two families of curves that are locally orthogonal passing through a point. Rabinovich et al. presented a concept of “discrete orthogonal geodesic nets,” which can realize consistent deformation of developable surfaces \cite{rabinovich_discrete_2018}. After that, they also developed the method for deformation of discrete surfaces with curve constrain \cite{rabinovich_shape_2018} and further developed an interactive design method \cite{rabinovich_modeling_2019}. Ion et al. approximated curved geometries with piece-wise developable surfaces also based on “discrete orthogonal geodesic nets” \cite{ion_shape_2020}. Wang et al. proposed a method of “discrete geodesic parallel coordinates,” that is, one family is geodesic and another is orthogonal to the geodesic family at least locally \cite{wang_discrete_2019} (see Fig. \ref{geodesic}b).

Distance field is another constructive method of the geodesic net formed by equidistant offsetting geodesic lines on the target surface starting from a point. One application of this method is the formation of deployable grids, as demonstrated in the work by Pillwein et al. \cite{pillwein_elastic_2020}, and deployable grids with the structure of non-convex hull are further discussed \cite{pillwein_generalized_2021}. The distance field technique is also utilized in generating knitting paths on surfaces \cite{liu_knitting_2021}. This approach allows for the direct production of non-developable surfaces without the need for cutting and stitching operations (see Fig. \ref{geodesic}c), and provides some ideas for the research of 3D knitting path generation in recent years (see fig. \ref{fabrication}e).

\textbf{Spanning Tree.}
In graph theory, a spanning tree is a connected subgraph that contains all the nodes of the graph but with the least number of edges. As the mesh can be transformed into a dual mesh that a graph, namely, a mesh face $f_{i} \in F$, as a node, is connected to the adjacent mesh faces $f_{j} \in \mathcal{N}(i)$ thus forming corresponding edges, a spanning tree can be used to represent the connectivity of the mesh faces, with unconnected nodes becoming segmentation lines of the mesh. Thus, the problem of mesh segmentation can be transformed into a problem of connection of nodes.

\begin{figure*}[t!]
    \centering
    \includegraphics[width=\textwidth]{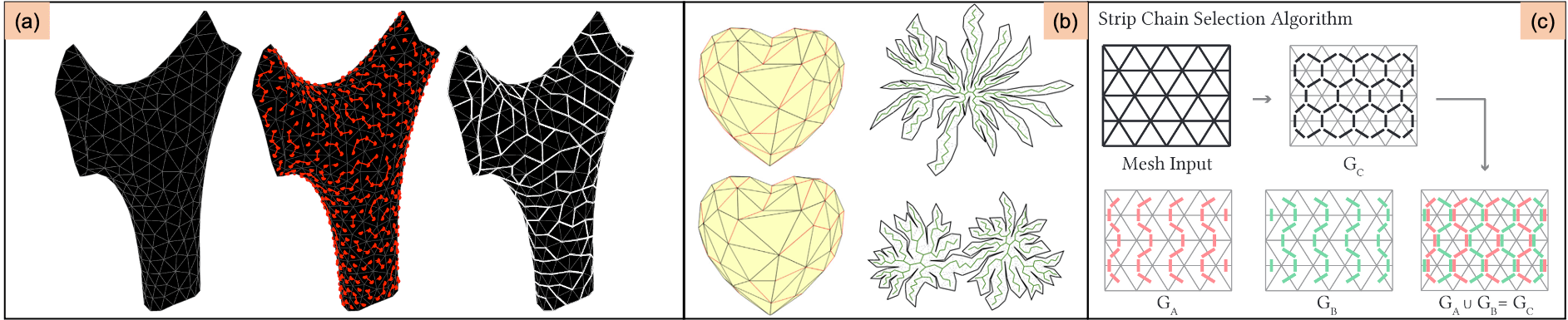}
    \vspace{-20pt}
    \caption{Spanning tree: (a) Greedy algorithms for minimum spanning Tree \cite{chandra_computing_2015}. (b) Segmentation and unfolding of Polygonal shapes by minimum spanning tree \cite{xi_learning_2016}. (c) Strip chains generation based on spanning tree \cite{leung_prototyping_2018}.}
    \label{tree}
\end{figure*}

Some methods of spanning trees are available for non-regular mesh. Basing on greedy algorithms, Chandra et al. obtained minimum spanning tree \cite{chandra_computing_2015} (see Fig. \ref{tree}a). Xi et al. grouped nodes by clustering to generate a spanning tree \cite{xi_learning_2016} (see Fig. \ref{tree}b). Kim, Xi, and Lien extended a genetic-based algorithm to optimize a spanning tree of the mesh of convex shell \cite{kim_disjoint_2017}. For the regular mesh, some studies use the connection relation of the four-edge data structure to generate intersecting assembly structures \cite{peraza_hernandez_smi_2013, akleman_construction_2016}. Leung generated long chains by entering regular meshes, which are used to weave into surfaces \cite{leung_prototyping_2018} (see Fig. \ref{tree}c).

\begin{figure*}[t!]
    \centering
    \includegraphics[width=\textwidth]{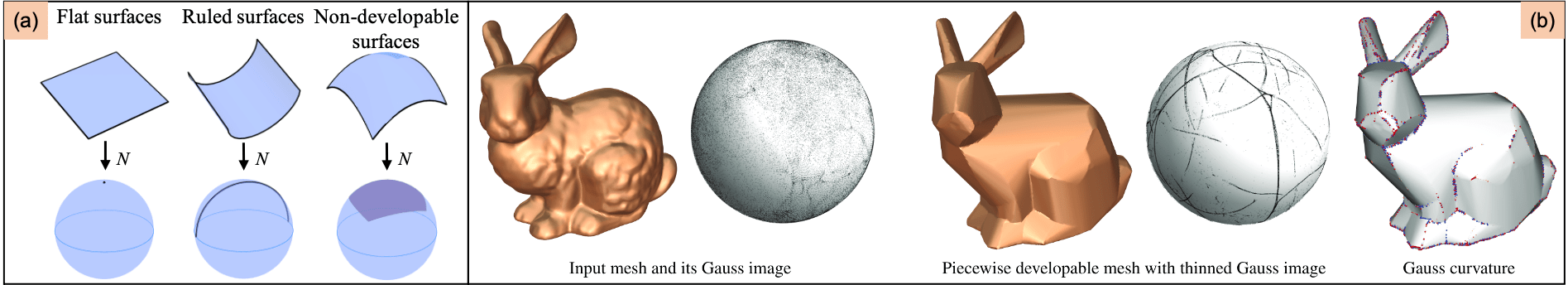}
    \vspace{-20pt}
    \caption{Gauss map: (a) The Gauss images of different surfaces. (b) Seam generation by Gauss map \cite{binninger_developable_2021}.}
    \label{gauss}
\end{figure*}

\textbf{Gauss Map.}
Gauss map refers to the normals of a 3D shape mapped to a sphere of radius one, and the local flatness of the target shape can be represented by the distribution of surface points on the unit sphere, that is, Gauss image (see Fig. \ref{gauss}a). The two characteristics of developable surfaces by Gauss map are as follows: for a flat surface, only a point is on the sphere because every normal on the surface is the same; for ruled surfaces, curves are presented on the sphere by Gauss map because each normal on a ruled surface rotates around an axis. Thus, undevelopable surfaces represent scattered points on the sphere by Gauss map. The method of segmentation based on the Gauss map is essentially the aggregation of scattered points on the sphere into a few points or curves, which converts the target shape into a shape composed of some developable surfaces, so dividing lines on the target shape form automatically. Available techniques for aggregation of scattered points include clustering algorithm for discrete Gauss map \cite{tang_optimization_2005} and computing minimum energy for Gauss map \cite{zhao_developability-driven_2022}.

Some studies are on the aggregation of scattered points on a unit sphere into curves. A free-form surface is approximated into a quasi-developable mesh by fitting a curve based on discrete Gauss map \cite{zeng_least_2012}. CAD model is divided into some parts \cite{zeng_developable_2021}, and a method of projecting points onto a curve realizes Gauss image thinning \cite{binninger_developable_2021} (see Fig. \ref{gauss}b). Some studies are on the aggregation of scattered points on the unit sphere into several points. A method of narrowing down the scope of “vertices star” is proposed, namely, narrowing down the scope of the region means flattening the mesh \cite{stein_developability_2018}. A method of narrowing down the scope is also proposed, which involves calculating a degenerate point from the minimum distance between points on a Gauss image and a cutting plane \cite{gavriil_optimizing_2019}.

\textbf{Vector Field.}
A vector field can be represented by vectors located at each point on a surface, which is a 2D manifold. For instance, a vector field can be constructed by tangent vectors at each vertex of a mesh, where each position has a direction and length. Vector fields are often generated based on geometric properties such as principal curvature and geodesics. The types of vector fields are many, and Vaxman et al. reviewed the different directional fields \cite{vaxman_directional_2016}. A network can be constructed according to the guidance of a vector field, which is a method of re-meshing.

Regarding re-meshing by a vector field, Sageman-Furnas et al. used a vector field to construct a global discrete “Chebyshev network” (all edge lengths of quad mesh are equal) to approximate a 3D shape \cite{sageman-furnas_chebyshev_2019}. Jiang et al. arranged the panels on a surface according to the “killing vector field,” which means a vector field that preserves measurements \cite{jiang_using_2021}. He et al. constructed a vector field by minimum principal curvatures, which are tangent vectors on all the vertices of a mesh. They then generated developable surfaces by flank milling tool path \cite{he_quasi-developable_2021}. Verhoeven et al. constructed the vector field through the direction of principal curvature, thus aligning the vector field with the re-meshing meshes \cite{verhoeven_dev2pq_2022}. In addition, cross-field means 4-rotational symmetry vector field, which is also used for re-meshing and segmentation, such as re-meshing quad network by cross-field realizes masonry of 3D shell in architecture \cite{panozzo_designing_2013}. A re-meshing method of cross-field is used to align sharp characteristics of a given 3D shape \cite{zhang_octahedral_2020}. A 3D garment segmentation is put forward under the guidance of user-interaction based on cross-field \cite{pietroni_computational_2022} (see Fig. \ref{interactive}a), which users divide a garment by setting of start points.


\textbf{Curvature.}
Curvature is a description of the bending condition of a curve. It is extended to surfaces. As a point on a surface can pass through many curves that are formed by the intersection of the tangent planes of the normal vector located at the point and the surface, the curvatures at a point on the surface are many. They are called normal curvatures. The maximum and minimum normal curvatures are called principal curvatures, and they are perpendicular to each other. Two types of curvature are formed based on principal curvature. Mean curvature is the average of the two principal curvatures, which locally describes the curvature of a surface. Gaussian curvature is the product of the two principal curvatures and can be used to determine whether a surface is developable or not. Curvature is an important geometric concept for developable surfaces, many studies realize segmentation according to curvature, which makes each part as easy to be flattened as possible.

\begin{figure}[ht]
    \centering
    \includegraphics[width=\linewidth]{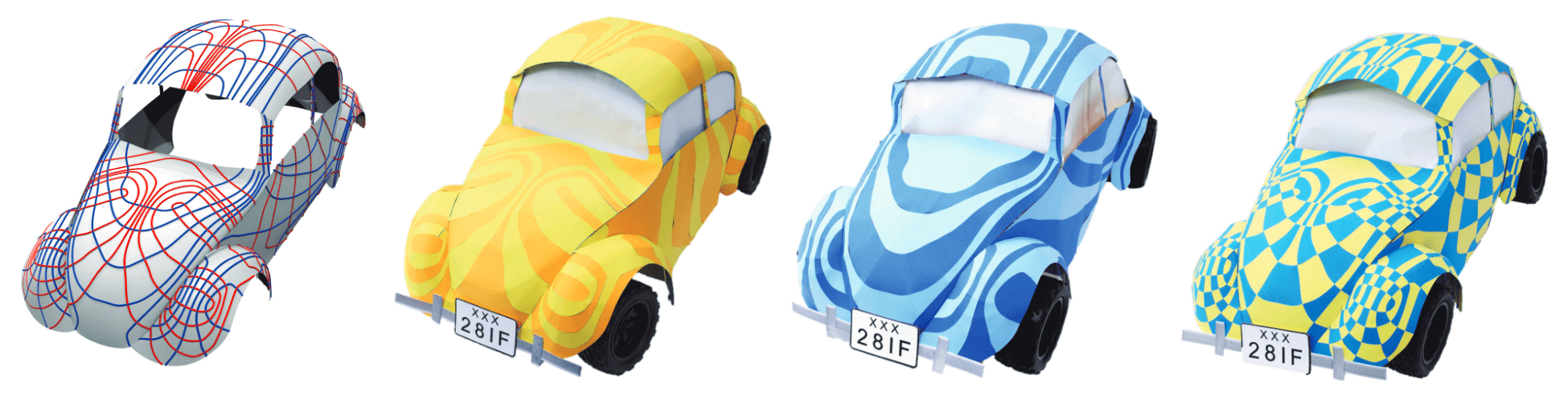}
    \vspace{-20pt}
    \caption{Construction of principal strips for woven 3D shapes by principal curvature \cite{takezawa_fabrication_2016}}
    \label{principal}
\end{figure}

For principal curvature, some studies generate orthogonal nets on curved surfaces according to the properties of principal curvature. For instance, a 3D shape is divided into strips through orthogonal curvature lines, which are formed by the principal curvature \cite{takezawa_fabrication_2016}. Ruled surfaces are reconstructed by the constraint that the principal curvature is aligned to the ruled lines of the surface \cite{wakamatsu_virtual_2017} (see Fig. \ref{principal}). 3D shapes are re-meshed into quad mesh by principal curvature \cite{pellis_principal_2020}, and shapes are re-meshed into quad meshes through vector fields constructed by the principal curvature direction \cite{verhoeven_dev2pq_2022}.
For Gaussian curvature, surfaces are divided by Gaussian curvature of zero \cite{chen_g2_2013} and vanishing Gaussian curvature in Gauss image \cite{zeng_least_2012, bhooshan_applying_2015, stein_developability_2018, gavriil_optimizing_2019, zeng_developable_2021, binninger_developable_2021, zhao_developability-driven_2022}. 
For geodesic curvature, it describes the curved condition of the curve embedded in a surface, that is, it describes the bending of deviating from the geodesic. Joo et al. propose a method of subdividing surfaces by constructing curvature lines based on geodesic curvature \cite{joo_differential_2014}.

\textbf{Origami.}
Origami is the shaping from 2D to 3D, and the term refers to more than just this art. Origami involves many geometric problems, so computational origami has become an important research field. The study of the return of 3D forms to a flat surface inspired by origami is also a segmentation, which focuses on generating creases.

Given a target polygonal shape, corresponding creases are generated based on the mesh faces of the shape as well as the corresponding connecting faces to assist in flattening. \cite{tachi_origamizing_2010, jiang_freeform_2020} (see Fig. \ref{tuck}). The polygonal shape is formed from a plane by folding, and the excess part is hidden inside the mesh. This method is also extended to 3D printing on a stretchable material \cite{guseinov_curveups_2017}. Curved creases are also generated based on quad mesh \cite{jiang_curve-pleated_2019}.

\textbf{Custom.}
Custom segmentation requires the user to provide specific operations with interaction, so operators can handle 3D shapes according to their ideas and experiences. From collected studies, custom segmentation operations can be divided into five categories: cross-section-based, loft-based, feature-based, rule-based, and experience-based.

\begin{figure*}[t!]
    \centering
    \includegraphics[width=\textwidth]{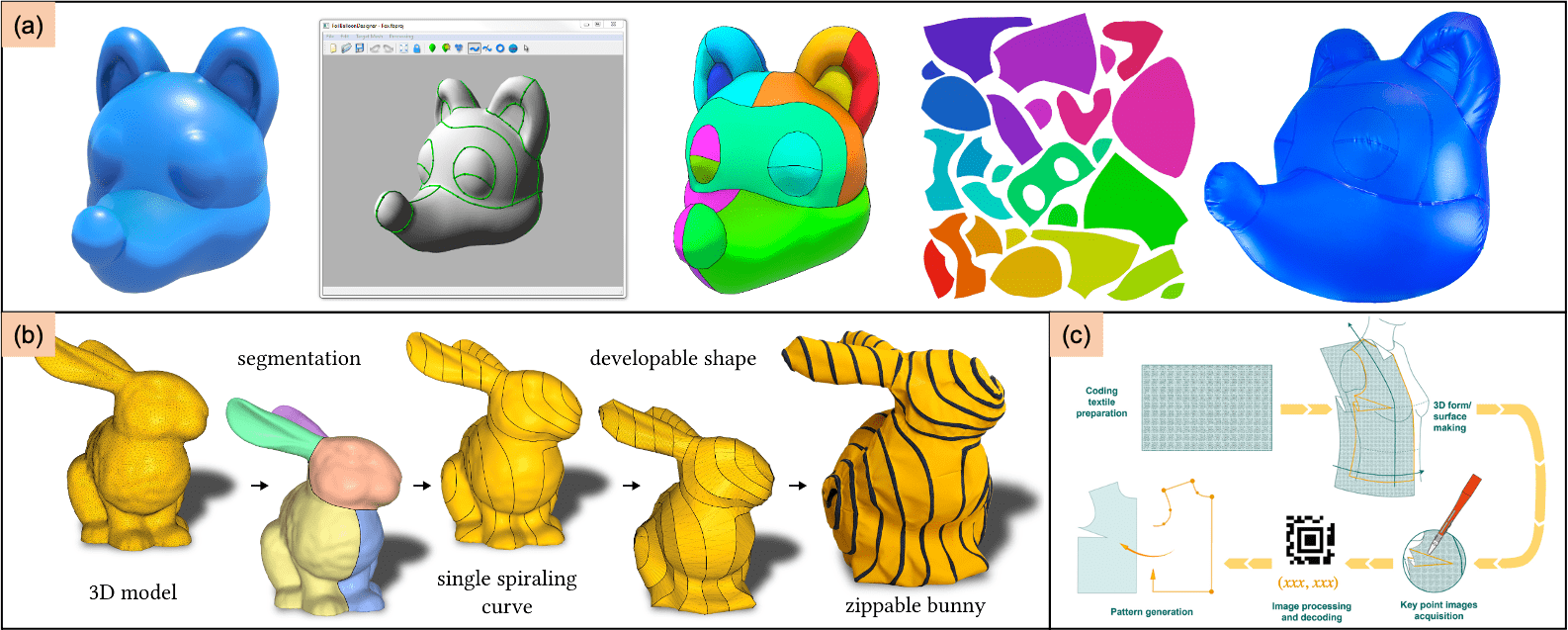}
    \vspace{-20pt}
    \caption{Custom: (a) 3D shapes segmentation according to some obvious geometric features \cite{lin_making_2014}, (b) Segmentation of closed shapes by spiral curves \cite{schuller_shape_2018}. (c) Segmentation according to experience in 3D garment design \cite{lei_new_2022}.}
    \label{custom}
\end{figure*}

Cross-section-based means horizontal segmentation according to the cross-section of 3D shapes that are columnar. For instance, 3D garments are split according to the characteristics of human cross-sections \cite{huang_block_2012}, and a vase unfolds by cross sections segmentation \cite{liu_automatically_2017}. Loft-based means the generation of developable surfaces according to two spatial curves, mainly used to shape generation based on loft curves \cite{liu_industrial_2011, chen_g2_2013, chen_technical_2013, bo_multi-strip_2019}. Feature-based means that shape segmentation is according to some obvious geometric features \cite{lin_making_2014, skouras_designing_2014, zhang_computational_2019} (see Fig. \ref{custom}a). Rule-based means splitting 3D shapes into special pieces according to specific rules, which often produces interesting creativity. For instance, a method is proposed to create shapes of the Mosaic effect using lots of different non-convex hulls based on a principle of attraction and repulsion \cite{chen_fabricable_2017}. A segmentation method of spiral curves is employed for unfolding closed shapes \cite{schuller_shape_2018} (see Fig. \ref{custom}b). Experience-based means that shapes segmentation according to experience in some specific design fields, such as interactive 3D garment design \cite{suh_optimal_2021, lei_new_2022} (see Fig. \ref{custom}c).

Subdivision is a method of discrediting target surfaces directly into a lot of small and simple meshes. Geodesic is a method of splitting surfaces based on local minimum distance, which often generates strip shapes or reconstruct mesh. Origami is a special method of fitting target shapes by folding. Gauss map is a method of degrading target surfaces into developable surfaces in Gauss image, but the shape of the segmentation is not controllable. The vector field is a method to guide the generation of curves on a surface, and manually defined start points are usually required, which is often used for interactive segmentation. Curvature is a method of describing the bending of curves or surfaces. Principal curvature is often used to split the target surface to form an orthogonal grid. Custom is a method of segmentation by manual drawing or custom segmentation, but custom exists the challenge of operating experience of shapes with different geometries. 

We also collect some studies (2015–2022) on AI-based mesh segmentation, which is not employed in the field of developable surfaces. Mesh segmentation is foundational research in the field of computer vision and computer graphics. A survey review traditional methods of mesh segmentation \cite{theologou_comprehensive_2015}. With the development of AI technology, new ways are established. Some research ideas are to provide the underlying topological mesh through mesh simplification, which is beneficial for subsequent mesh segmentation and mesh parameterization, e.g., “QEM-based mesh simplification” \cite{ozaki_out--core_2015} and “MeshCNN” \cite{hanocka_meshcnn_2019}. Research also implements semantic-based shape segmentation by training mesh labels \cite{guo_3d_2016, maron_convolutional_2017, herouane_labelling_2018, george_deep_2022}. Other studies are on shape segmentation by learning geometric features of the mesh, e.g., “geodesic neural network” \cite{poulenard_multi-directional_2018}, “CurvaNet” \cite{he_curvanet_2020}, graph neural network by mesh dual \cite{milano_primal-dual_2020, wen_dual-sampling_2021}, recurrent neural network by random walks on mesh \cite{lahav_meshwalker_2020}, “subdivision-based CNN” \cite{hu_subdivision-based_2022}, and diffusion model based on mesh segmentation \cite{sharp_diffusionnet_2022}. from these studies, the segmentation of developable surfaces can be inspired by supervised learning methods.

In general, existing segmentation methods can deal with most shapes, but for different specific tasks, custom segmentation methods are still very necessary. Some research focuses on interaction, which is essentially the use of human experience. Further, by learning a large amount of manual segmentation data, AI can generate segmentation results more in line with requirements.

\subsection{Surface Flattening}\label{sec3.2}

This section focuses on flattening free-form surfaces into developable surfaces. Surface flattening aims to minimize errors between before and after deformation. During deformation, it is essential to preserve certain geometric properties as much as possible. Four types of mesh properties can be maintained before and after deformation: \textbf{edge preserving}, \textbf{position preserving}, \textbf{angle preserving}, and \textbf{area preserving}. Consequently, surface flattening can be approached as a numerical optimization problem that aims to minimize energy. In order to achieve specific design objectives (as shown in Table \ref{optimal}), several studies have employed multi-mesh property preservation to construct energy functions that minimize distortion.

\begin{table*}[t!]
    \centering
    \caption{Achievement of different design objectives by corresponding optimization functions.}
    \label{optimal}
    \vspace{-5pt}
    \begin{tabular}{|l|c|c|c|c|r|}
         \hline
         Design goals & $E_{ang}$ & $E_{area}$ & $E_{pos}$ & $E_{edge}$ & Examples \\ \hline
         Preserve flatness of mesh & \cellcolor[gray]{0.8} & & & & \cite{bhooshan_applying_2015} \\ \hline
         Preserve corners of mesh & \cellcolor[gray]{0.8} & & & & \cite{sasaki_simple_2022} \\ \hline
         Set anchor points & & & \cellcolor[gray]{0.8} & & \cite{zhang_computational_2019} \\ \hline
         Projection & & & \cellcolor[gray]{0.8} & & \cite{ren_3d_2021} \\ \hline
         Isometric mapping & & & & \cellcolor[gray]{0.8} & \cite{jiang_using_2021} \\ \hline
         Edge alignment & & & & \cellcolor[gray]{0.8} & \cite{guseinov_curveups_2017} \\ \hline
         Construction of orthogonal net & \cellcolor[gray]{0.8} & & \cellcolor[gray]{0.8} & & \cite{takezawa_fabrication_2016} \\ \hline
         Shape fitting by planar texture & \cellcolor[gray]{0.8} & & \cellcolor[gray]{0.8} & & \cite{jiang_polyhedral_2015} \\ \hline
         Simulation of paper bending & \cellcolor[gray]{0.8} & \cellcolor[gray]{0.8} & \cellcolor[gray]{0.8} & & \cite{signer_developable_2021} \\ \hline
         Fabric deformation & & \cellcolor[gray]{0.8} & \cellcolor[gray]{0.8} & \cellcolor[gray]{0.8} & \cite{zhu_research_2022} \\ \hline
         Simulation of "Auxectic" structure & & & \cellcolor[gray]{0.8} & \cellcolor[gray]{0.8} & \cite{zhu_research_2022} \\ \hline
         Deformation of stretchable fabric & & & \cellcolor[gray]{0.8} & \cellcolor[gray]{0.8} & \cite{zhang_computational_2019} \\ \hline
         Shape fitting by developable surface & & & \cellcolor[gray]{0.8} & \cellcolor[gray]{0.8} & \cite{jiang_freeform_2020} \\ \hline
    \end{tabular}
\end{table*}

\textbf{Edge Preserving.}
Edge preserving means keeping the length of mesh edges as constant as possible during the mesh deformation. For rigid transformation, each part of the shape is only translated or rotated during deformation. Owing to the stability of the triangular mesh, the shape can be maintained as long as each edge of the mesh is kept as much as possible by rigid transformations. Thus, the minimum energy function is established by optimizing the edge lengths before and after deformation \cite{sorkine2007rigid}, which can be expressed as follows:

\begin{equation}
    E_{edge}(p^{\prime}, R)=\arg\min\sum_{i}\sum_{j \in \mathcal{N}(i)} w_{i j}||(p_{i}^{\prime}-p_{j}^{\prime})-R_{i}(p_{i}-p_{j})||^{2}
\end{equation}
where $\mathcal{N}(i)$ is the set of vertices adjacent to $p_{i}$, $p_{i}$ and $p_{i}^{\prime}$ are the vertices before and after the deformation, $R$ are the rotation matrices. The weights $w_{ij}$ are the average or cotangent weights.

Edge preserving is often used to maintain shape and simulation of elastic deformation. For instance, Konakovic et al. implemented isometric mapping by rigidity constraint which is the minimum energy function by the least squares method with the projection length and the output length \cite{konakovic_beyond_2016}. Rabinovich et al. constructed isometric constrain with the length in a reference mesh \cite{rabinovich_shape_2018}. Similar studies are conducted \cite{zhang_computational_2019, rabinovich_modeling_2019, zhao_developability-driven_2022}. In addition, singular values of $\sigma_{1}$ and $\sigma_{2}$ of the Jacobian matrix are metric distortion, and if the following equation is satisfied:
\begin{equation}
    \sigma_{1}=\sigma_{2}=1
\end{equation}
which is an isometric mapping, where the Jacobian matrix is equivalent to a rigid rotation matrix.
Isometric distortion with the singular values as independent variables is constructed as follows:
\begin{equation}
    E_{iso}=\sum_{t}((\sigma_{1}-1)^{2}+(\sigma_{2}-1)^{2})
\end{equation}
The energy function is minimized to approximate the rigid rotation matrix \cite{liu_computational_2019, schuller_shape_2018}.

In addition to implementing isometric mapping, some geometric constraints must be satisfied depending on the specific requirements, such as the position of certain vertices being set or the angle of the mesh being fixed as follows:
\begin{equation}
    p_{k}=\hat{p}_{k}, k \in M
\end{equation}
where $M$ is the set consisting of discrete points at constrained locations, $p_{k}$ are the vertices of the output mesh, and $\hat{p}_{k}$ are the constraint positions by the user. However, to minimize the energy function as well as satisfy the constraint, constraints can also be transformed into energy functions, such as position preserving, angle preserving, and area preserving.

\textbf{Position Preserving.}
Position preservation refers to the preservation of mesh vertices' positions as closely as possible during mesh deformation. This property is commonly utilized to deform the mesh and fit it onto a target surface or to establish anchor points. For example, mesh fitting onto a target surface can be achieved through point projection techniques \cite{konakovic_beyond_2016, jiang_quad-mesh_2020, jiang_freeform_2020}, while constrained positions can be used to set anchor points \cite{rabinovich_shape_2018, zhao_developability-driven_2022}. The minimum energy function associated with position preservation can be expressed as follows:
\begin{equation}
    E_{pos}(p_{k}) = \arg\min\sum_{k \in M}||p_{k}-\hat{p}_{k}||^2
\end{equation}

\textbf{Angle Preserving.}
Angle preserving is keeping the angle as constant as possible during the mesh deformation. The minimum energy function can be expressed as:
\begin{equation}
    E_{ang}(\theta_{i}) = \arg\min\sum_{i} \omega_{i}(\theta_{i} - \alpha_{i})^2
\end{equation}
where $\omega_{i}$ are the weights of the angle, which usually corresponds to the edge length of the corner; $\theta_{i}$ are the angle of output mesh faces; and $\alpha_{i}$ are the angle of input mesh faces or the setting angle of the goal.

For instance, angle preserving is usually used to maintain the grid as orthogonal \cite{takezawa_fabrication_2016, sasaki_simple_2022}, maintains parallelism between adjacent faces \cite{signer_developable_2021}, simulate materials folding \cite{peraza_hernandez_modeling_2016} and simulate force flattening surfaces \cite{bhooshan_applying_2015}.
In addition, angles between vectors are constrained to make the mesh faces perpendicular to the surface normals \cite{jiang_polyhedral_2015, jiang_curve-pleated_2019, he_quasi-developable_2021}. Adjacent normals are aligned to deform the surface into developable surfaces in Gauss map \cite{stein_developability_2018}. The angle between vectors is constrained to reconstruct the mesh \cite{sageman-furnas_chebyshev_2019, verhoeven_dev2pq_2022}.

\textbf{Area Preserving.}
Area preserving means keeping the area of mesh faces as constant as possible during the mesh deformation, which can be used for the simulation of the deformation of elastic fabric \cite{zhu_research_2022} folding of paper \cite{rabinovich_shape_2018}. The minimum energy function can be expressed as follows:
\begin{equation}
    E_{area}(a_{i}) = \arg\min\sum_{i} \omega_{i}(a_{i} - \hat{a}_{i})^2
\end{equation}
where, $a_{i}$ are the area of the output mesh faces, and $\hat{a_{i}}$ are corresponding faces area before deformation. $\omega_{i}$ are usually related to the size of the $i$th mesh face.

In conclusion, the energy functions constructed from these geometric metrics can be used to constrain specific geometric properties while flattening surfaces. Thus, a total energy function can be expressed as follows:
\begin{equation}
    E = \omega_{ang}E_{ang}+\omega_{area}E_{area}+\omega_{pos}E_{pos}+\omega_{edge}E_{edge}
\end{equation}
The minimization of energy becomes a numerical solution problem, which allows the constrained optimization problem to be transformed into an unconstrained optimization problem and easily computed. Several common optimization algorithms can be used to solve the question, such as least squares, gradient descent, Newton method, and so on. During iterative calculation, the result can be obtained when the value is less than the error or exceeds the maximum number of iterations.

\section{Physical Modeling}\label{sec4}

Physical modeling involves the production of digital models in physical form. The objective of Section \ref{sec3} is to generate either a planar structure as a whole or multiple planar parts that can be assembled to create a 3D shape. Consequently, this section explores different approaches for the \textbf{manufacturing} and \textbf{assembly} of physical models representing 3D shapes (see Table \ref{physical_modeling}). The availability of various unfolding methods and materials allows for different ways of realizing physical models.

\begin{figure*}[t!]
    \centering
    \includegraphics[width=\textwidth]{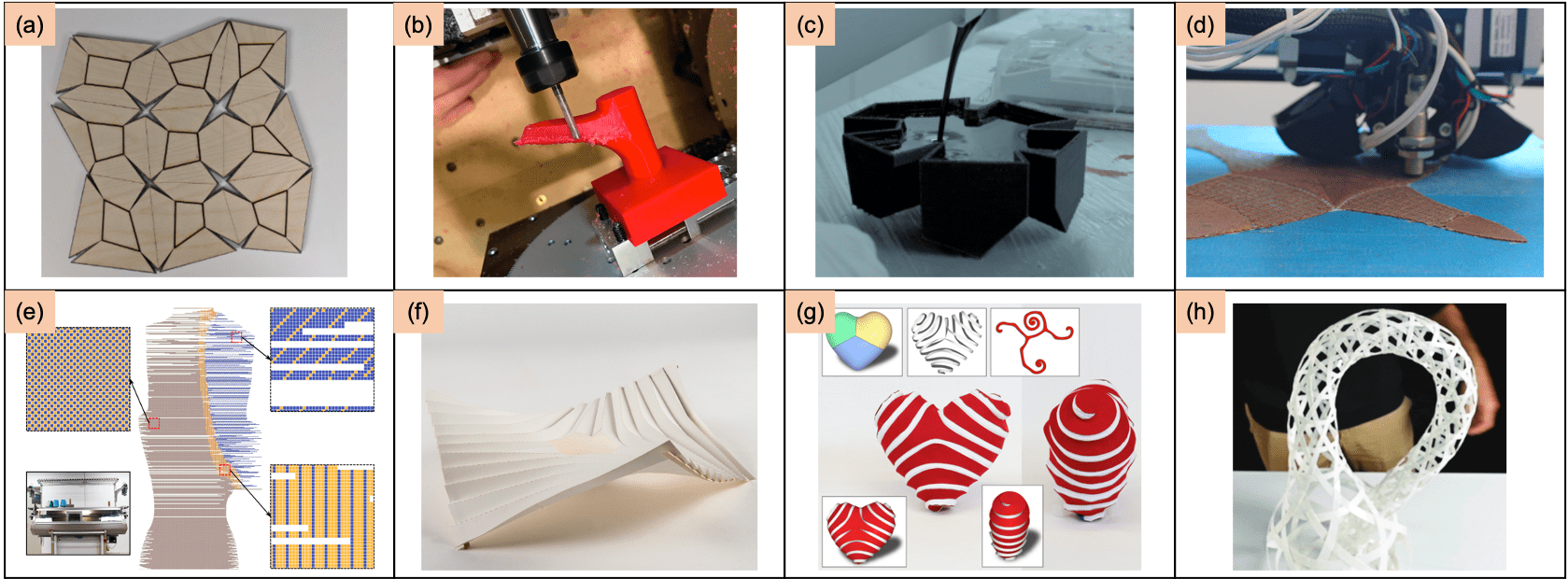}
    \vspace{-20pt}
    \caption{Selected samples of physical modeling: (a-e) are figures of manufacturing and (f-h) are figures of assembly. (a) Cutting \cite{jiang_freeform_2020}, (b) Milling \cite{stein_developability_2018}, (c) Casting \cite{ erdine_curved-crease_2019}, (d) 3D-printing \cite{noma_pop-up_2020}, (e) Knitting \cite{liu_knitting_2021}, (f) Folding \cite{jiang_curve-pleated_2019}, (g) Joint \cite{schuller_shape_2018}, (h) Woven \cite{ren_3d_2021}.}
    \label{fabrication}
\end{figure*}

\begin{table*}[t!]
    \centering
    \caption{Features of different physical forming methods.}
    \label{physical_modeling}
    \vspace{-5pt}
    \begin{tabular}{|c|c|c|c|}
         \hline
         Physical Modeling & Merits & Defects & Examples \\ \hline
         Cutting & \makecell[c]{Fast speed of production \\ Low cost of production \\ Wide selection of materials} & \makecell[c]{Difficult to assemble \\ Unavoidable error} & \cite{akleman_construction_2016} \\ \hline
         Milling & \makecell[c]{Suitable for producing complex surfaces \\ Wide selection of materials \\ High machining accuracy} & \makecell[c]{High cost of production \\ Difficult to produce large scale shapes} & \makecell[c]{\cite{he_quasi-developable_2021} \\ \cite{stein_developability_2018}} \\ \hline
         Casting & \makecell[c]{Fast speed of production \\ Suitable for producing complex surfaces \\ Not limited by shape thickness} & \makecell[c]{More work procedures \\ Poor mechanical properties of shapes} & \makecell[c]{\cite{skouras_designing_2014} \\ \cite{zhang_computational_2019}} \\ \hline
         3D-Printing & \makecell[c]{Suitable for producing complex surfaces \\ Suitable for mass custom production} & \makecell[c]{Limited choice of materials \\ Poor mechanical properties of shapes} & \makecell[c]{\cite{chen_fabricable_2017} \\ \cite{tahouni_self-shaping_2020}} \\ \hline
         Knitting & \makecell[c]{Fast speed of production \\ Low cost of production \\ Suitable for producing flexible materials} & \makecell[c]{Limited choice of materials \\ Unavoidable error} & \makecell[c]{\cite{schuller_shape_2018} \\ \cite{liu_knitting_2021}} \\ \hline
         Folding & Simple processing & Unavoidable error & \cite{noma_pop-up_2020} \\ \hline
         Joint & Strong connection & Complex assembly process & \cite{akleman_construction_2016} \\ \hline
         Woven & Not require additional fixation & \makecell[c]{Complex assembly process \\ Unavoidable error} & \cite{ren_3d_2021} \\ \hline
    \end{tabular}
\end{table*}

\subsection{Manufacturing}\label{sec4.1}

Manufacturing processes can be classified into two main approaches based on the material form: subtractive manufacturing and additive manufacturing. Subtractive manufacturing involves reducing materials from their original entirety to produce 3D shapes. On the other hand, additive manufacturing involves adding materials to create 3D shapes. Subtractive manufacturing offers advantages in terms of fabrication efficiency and material selectivity. However, it has limitations when it comes to producing 3D shapes with complex and free-form surfaces. In contrast, additive manufacturing is well-suited for creating 3D shapes with intricate free-form surfaces. However, it may have limitations in terms of material selectivity.

According to collected studies where the type of manufacturing is used, subtractive manufacturing can be subdivided into two ways, which are \textbf{cutting} and \textbf{milling}. Additive manufacturing can be subdivided into three details ways, which are \textbf{casting}, \textbf{3D-printing} and \textbf{knitting}.

\textbf{Cutting.}
Cutting is one of the most common manufacturing techniques used in the forming of planar materials. Cutting means  cutting along specified paths on a flat material to form flat patterns. The feature of the cutting method is the use of thin and flat materials. Cutting, is one of the flat manufacturing methods with many merits: it is fast and inexpensive to produce and can be made from a wide selection of materials as long as they are flat. For its defects, the assembly of 3D shapes is challenging because it usually consists of multiple planar parts, and unavoidable errors occur between real and digital shapes due to the planar approximation of surfaces.

Prototyping by paper cutting are discussed in some studies, such as \cite{peraza_hernandez_smi_2013, lin_making_2014, akleman_construction_2016, xi_learning_2016, takezawa_fabrication_2016, vergauwen_computational_2017, kim_disjoint_2017, leung_prototyping_2018, jiang_freeform_2020, zhao_developability-driven_2022} (see Fig. \ref{fabrication}a).
Fabrication by metal cutting is discussed in \cite{sevtsuk_freeform_2014, bhooshan_applying_2015, konakovic_beyond_2016}; fabrication by hard material cutting, \cite{konakovic-lukovic_rapid_2018}; cloth cutting, \cite{wang_pattern_2010, huang_block_2012, zeng_least_2012, wakamatsu_virtual_2017, schuller_shape_2018, zhang_computational_2019}.

\textbf{Milling.}
Milling is one of the mechanical methods that uses a milling cutter as a tool to machine the surface of an object. The feature of Milling is removing excess material following a defined path by a high-speed rotating tool head. Developable surfaces are generated by path planning to control the action of the tool head, thus the 3D shape can be fitted by developable surfaces and then manufactured by milling, such as \cite{stein_developability_2018, he_quasi-developable_2021} (see Fig. \ref{fabrication}b). On the one hand, milling has some merits. Milling can produce free-form surfaces, and it has a high machining accuracy due to the high degree of freedom of the tool head. In addition, many materials can be selected. On the other hand, the disadvantages of milling are the high cost of production and the difficulty to produce large-scale shapes due to the limitation of machines.

\textbf{Casting.}
Casting is to form a shape through a specific mold. Casting is used to production of the panels in building \cite{eigensatz_paneling_2010, fu_k-set_2010, singh_triangle_2010, bo_circular_2011, yang_shape_2011, panozzo_designing_2013, erdine_curved-crease_2019} (see Fig. \ref{fabrication}c), and production of complicated shapes \cite{skouras_designing_2014, zhang_computational_2019}. On the one hand, Casting has some merits. The molding speed of 3D shapes is beneficial for mass production, and casting is suitable for making complicated shapes and is not limited by the thickness of shapes. On the other hand, more procedures need to be added due to the need to remove the mold, and general casting parts do not have good mechanical properties because the material is not compact inside.

\textbf{3D-Printing.}
3D printing technology is widely employed in research and prototyping due to its rapid molding speed and ability to produce complex surfaces with ease. It involves the gradual accumulation of materials to form a 3D shape. 3D printing is particularly advantageous when it comes to creating small yet intricate shapes. It allows for a seamless transformation from a digital model to a physical object by converting the digital model into a multi-layered planar path through G-code. Several studies have utilized this method, such as \cite{papanikolaou_bodypods_2015, peraza_hernandez_modeling_2016, chen_fabricable_2017, guseinov_curveups_2017, malomo_flexmaps_2018, noma_pop-up_2020, tahouni_self-shaping_2020} (see Fig. \ref{fabrication}d). 3D printing offers benefits in producing intricate surfaces and is suitable for mass customization. However, it has limitations, including a restricted range of materials, which primarily limits its use to the prototyping phase. Furthermore, researching the mechanical properties of the produced shapes remains a challenge.

\textbf{Knitting.}
Knitting technology is a manufacturing method that uses thread to form a 2D plane by knitting. General knitting methods produce flat fabrics \cite{wang_pattern_2010, huang_block_2012, zeng_least_2012, wakamatsu_virtual_2017, schuller_shape_2018, zhang_computational_2019}. These flat fabrics must be cut and sewn to form curved forms. The surface can be knitted directly through the knitting path that is generated by geodesic \cite{liu_knitting_2021} (see Fig. \ref{fabrication}e). The widespread use of CNC knitting machines makes production faster and cheaper, which is a method for making flexible materials. However, knitting is only suitable for thread-type materials and errors between real shapes and digital shapes are unavoidable.

\subsection{Assembly}\label{sec4.2}

The purpose of assembly is to reduce the difficulty of manufacturing. Some complex shapes can be made by breaking them down into simple parts. Therefore, assembling methods have been developed. Referring to the studies of collections, assembling can be subdivided into three types, which are \textbf{folding}, \textbf{joint}, and \textbf{woven} according to the results of segmentation and unfolding.

\textbf{Folding.}
Folding is a way to transform a plane into a free-form surface by setting creases with straight or curved lines. It does not require the combination of multiple parts and is essentially a deformation of a single shape. This approach is suitable for certain folding structures \cite{noma_pop-up_2020, tahouni_self-shaping_2020} (see Fig. \ref{fabrication}f). The merit of folding is the simplicity of the process; the defect is unavoidable errors.

\textbf{Joint.}
Joint is the combination of several parts into one whole by a specific connection. Depending on the material, different joints exist. Gluing or splicing can be used for paper, welding can be used for metal, and sewing can be used for soft material such as cloth. Other methods are zippable \cite{schuller_shape_2018} (see Fig. \ref{fabrication}g) and magnetic \cite{akleman_construction_2016}. The merit of the joint is that the two pieces have a strong connection, but the process of assembly is usually complex.

\textbf{Woven.}
Weaving is a method of forming a shape by interweaving many linear materials. Some studies use this method \cite{takezawa_fabrication_2016, leung_prototyping_2018, ren_3d_2021} (see Fig. \ref{fabrication}h). Weaving does not require additional fixation and has high stability, but the process of assembly is complex. Errors between real shapes and digital shapes are also unavoidable.

In conclusion, physical modeling is an important part of the transformation of a digital model into a real object, and the appropriate manufacturing and assembly methods can be selected according to different design results and material properties. Combining the advantages of developable surfaces with the advantages of fabrication, the "design to fabrication" method can be well used for rapid construction and the development of new materials, which is a promising research direction.

\section{Interactive Assistance}\label{sec5}

This section focuses on the relevant interactive assistance of digital modeling and relevant interactive approaches to developable surfaces. For digital modeling, interaction is effective in aiding shape segmentation and surface flattening. Some segmentation methods are not well suited to certain characteristics of the shape. For example, the shape of the generated parts is irregular, which makes production and assembly more difficult. More controllable results can be obtained with the aid of interactive segmentation. In surface flattening, many problems occur during the optimal computation, such as long iteration time and overlapping mesh faces. The global optimal solution is also difficult to obtain. In some cases, the direction of optimization can be imagined by visual observation, which can greatly reduce the number of iterations and allow for a more controlled and smooth unfolding by obtaining optimal solutions through interactive dragging and parameter adjustment. Interactive assistance can avoid some shortcomings of segmentation and flattening. In addition, relevant interactions are used to exchange information between digital and physical to obtain or enhance certain functions for design. According to our collection of studies, interactive assistance has two types. The first type is \textbf{unfolding assistant}, which is focused on the aid for shape segmentation and surface flattening. The second type is \textbf{tangible interaction}, which is focused on the communication between digital and physical.

\begin{figure*}[t!]
    \centering
    \includegraphics[width=\textwidth]{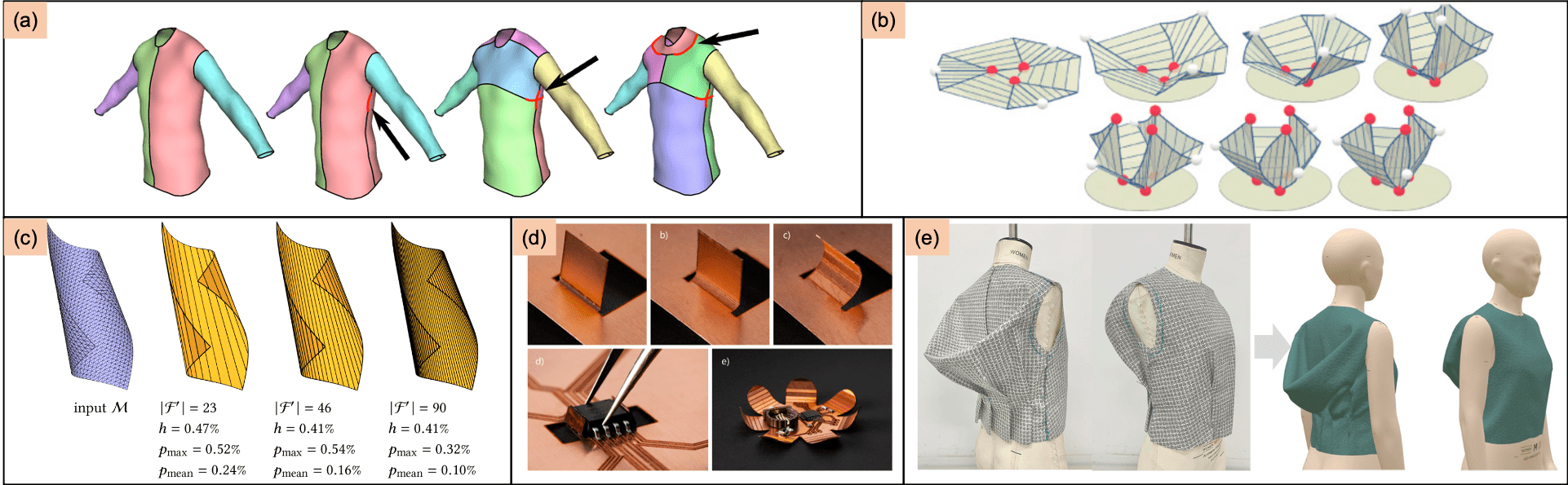}
    \vspace{-20pt}
    \caption{Selected examples of interaction: (a) Cutting for segmentation of clothes by guidance of users \cite{pietroni_computational_2022}. (b) Unfolding for surfaces by dragging \cite{tang_interactive_2016}. (c) Parameterizing for optimization by setting parameters \cite{verhoeven_dev2pq_2022}. (d) Numerical control for deformation of physical developable surfaces \cite{yan_fibercuit_2022}. (e) Physical control for generation of digital shapes \cite{lei_new_2022}.}
    \label{interactive}
\end{figure*}

\subsection{Unfolding Assistance}\label{sec5.1}

Unfolding assistance is suitable to assist in shape segmentation and surface flattening, which can be subdivided into three interactive operations, namely, \textbf{cutting} for custom drawing of curves on shapes, \textbf{unfolding} for dragging deformation, and \textbf{parameterizing} for control and adjustment.

\begin{wrapfigure}{l}{0.08\linewidth}
\includegraphics[width=1cm]{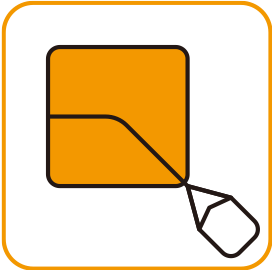}
\vspace{-20pt} 
\end{wrapfigure}

\noindent \textbf{Cutting.}
Cutting is the interactive process of dividing a shape by users drawing curves. For instance, developable surfaces are generated in real time by user-drawn curves on a digital board \cite{liu_industrial_2011}. A shape is divided into many surfaces by drawing feature lines on the original 3D shape for making sky lanterns \cite{lin_making_2014}. The user draws the structural frame lines of the shapes for segmentation \cite{skouras_designing_2014}, does shape subdivision \cite{gao_grid_2017}, cloth segmentation by user guidance \cite{pietroni_computational_2022} (see Fig. \ref{interactive}a), and draws creases in flat a surface for digital folding \cite{solomon_flexible_2012, paczkowski_paper3d_2014, vergauwen_computational_2017, kilian_string_2017, tahouni_self-shaping_2020, sasaki_simple_2022}. Cutting is mainly used for dividing a 3D shape into multiple surfaces. Some unfolding methods can automatically split 3D shapes, but the user-interactive approach is more appropriate for the control of details.

\begin{wrapfigure}{l}{0.08\linewidth}
\vspace{-10pt}
\includegraphics[width=1cm]{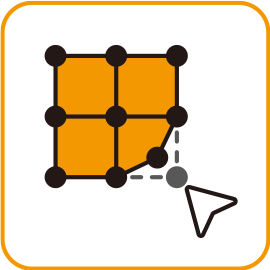}
\vspace{-20pt} 
\end{wrapfigure}

\noindent \textbf{Unfolding.} Unfolding involves the user dragging a digital model to accelerate surface flattening \cite{konakovic_beyond_2016} and construct folding structures through the graphical user interface. For instance, origami structure is generated by dragging the mesh faces \cite{tachi_origamizing_2010, tang_interactive_2016} (see Fig. \ref{interactive}b), Simulate flipping \cite{paczkowski_paper3d_2014}. Unfolding is a very free interactive operation to control the deformation of the surface, similar to deforming objects in the physical world.

\begin{wrapfigure}{l}{0.08\linewidth}
\vspace{-10pt}
\includegraphics[width=1cm]{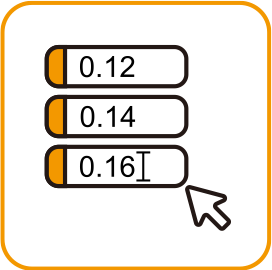}
\vspace{-20pt} 
\end{wrapfigure}

\noindent \textbf{Parameterizing.} In parameterizing, the user inputs or changes the relevant parameters to influence the numerical optimization and thus change the flattening results. This interactive operation precisely controls several parameters, leading to different levels of optimization \cite{binninger_developable_2021, verhoeven_dev2pq_2022, zhao_developability-driven_2022} (see Fig. \ref{interactive}c) or different types of optimization results that can be compared \cite{ion_shape_2020}. In addition, parameterizing is utilized to control and adjust CNC equipment (see Fig. \ref{fabrication}) used in physical modeling.

\subsection{Tangible Interaction}\label{sec5.2}

Tangible interaction is no longer limited to the interaction of the conventional computer interface; it extends the interaction to physical objects. It can form many new interactive ways wherein the information can be transferred between the physical objects and the computer. According to our collected studies, tangible interaction can be subdivided into two interactive operations, namely, \textbf{numerical control} and \textbf{physical control}.

\textbf{Numerical Control.} Numerical control refers to the controlling the deformation of physical shapes by computer. Given the deformable structural characteristics of the developable surfaces, studies are conducted on deformable numerical control robots \cite{lee_origami_2018} and programmable deformable structures \cite{purnendu_soft_2021, eguchi_pneumatic_2022, yan_fibercuit_2022, yang_compumat_2023} based on the inspiration of developable surfaces (see Fig. \ref{interactive}d).

\textbf{Physical Control.} Physical control refers to the user manipulating the physical object to make it morph in order to generate the corresponding digital signal to interact with the computer. Some shapes are difficult to digital simulation and require a lot of computational power, but they are easily shaped in the real world \cite{schreck_nonsmooth_2015, lei_new_2022} (see Fig. \ref{interactive}e). Based on the deformable property of developable surfaces, the relevant studies design tangible interfaces to implement some interactive functions, such as “bodyPod” \cite{papanikolaou_bodypods_2015}, “Kirigami Keyboard” \cite{chang_kirigami_2019}, “ExpanDial” \cite{kim_expandial_2019}, kirigami light \cite{zheng_sensing_2019}, and “Kirigami Haptic Swatches” \cite{chang_kirigami_2020}.

In conclusion, interactive assistance can aid shape segmentation and surface flattening, and most of the studies use interaction to simplify difficult tasks and make operation more controllable and fast. In addition, interactions can be extended to achieve new product functions through tangible interactions.

\section{Design Applications}\label{sec6}

This section focuses on applications based on relevant studies of developable surfaces. Depending on the field of application of the developable surface in the collected studies, we subdivide them into six application directions, which are \textbf{architecture}, \textbf{industrial products}, \textbf{arts and crafts}, \textbf{garments}, \textbf{structural materials} and \textbf{data physicalization}.

\begin{figure*}[t!]
    \centering
    \includegraphics[width=\textwidth]{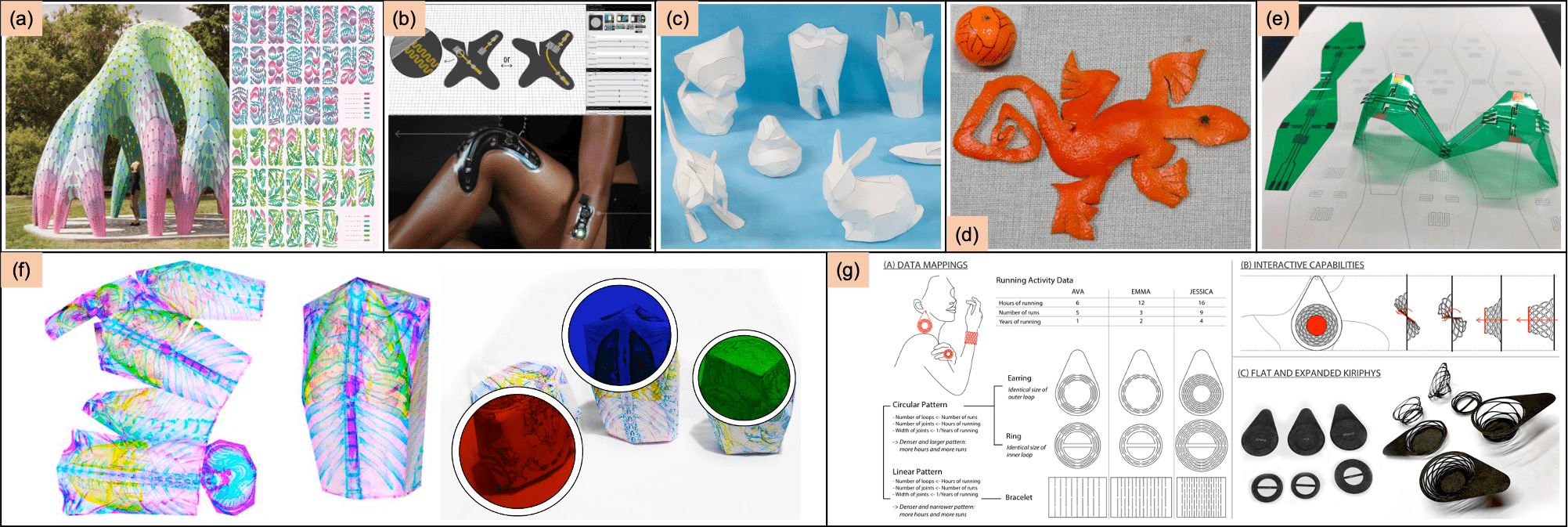}
    \vspace{-20pt}
    \caption{Selected samples of design applications: (a) shell construction \cite{marc_art_2016}, (b) wearable device \cite{markvicka_electrodermis_2019}, (c) origami \cite{zhao_developability-driven_2022}, (d) peeling art \cite{liu_computational_2019}, (e) origami robots \cite{lee_origami_2018}, (f) anatomical physical visualization \cite{schindler_anatomical_2020}, (g) jewelry design of data physicalization \cite{daneshzand_kiriphys_2022}.}
    \label{application}
\end{figure*}

\textbf{Architecture.}
Developable surfaces are widely used in architecture, where construction using component construction and assembly is necessary because of the extremely large spatial scale of buildings. complex curved buildings are constructed by discrete surfaces for a large number of planes or simple curved parts. According to the different applications of developable surfaces in curved buildings, the four application directions are facade, shell, bricklaying, and frame.
Facade is the exterior of a building, and the facade surface is subdivided into a large number of similar flat or simply curved panels, and the type of panels is reduced by employing methods such as clustering within less than the construction errors \cite{eigensatz_paneling_2010, fu_k-set_2010, singh_triangle_2010, yang_shape_2011, tang_form-finding_2014, gavriil_optimizing_2019}. The surface can also be transformed into a large number of strips of developable surfaces, and then the construction is completed through bending \cite{wang_discrete_2019}. These methods can approximate the surface of the building, but the assembly is difficult due to a large number of parts.
Shell is a self-supporting building form that is widely used due to its excellent force structure, space with large spans, and graceful curved shapes. Discrete meshes allow the construction of shells \cite{vouga_design_2012}. In addition, shells can be constructed using the combination of a large number of developable surfaces \cite{kahlert_width-bounded_2011, joo_differential_2014, gavriil_optimizing_2019, marc_art_2016} (see Fig. \ref{application}a). This approach allows more reduction of assembled parts compared with the discrete mesh approach.
Frame maintains the basic shape and support of the building, which are usually made by combining rod-like materials, and subdivision is an intuitive method of generating frames \cite{bo_circular_2011,jiang_freeform_2014,sevtsuk_freeform_2014,pottmann_cell_2015,jiang_planar_2022, bhooshan_applying_2015}. Frames can also be formed by combining long, elastic materials \cite{pillwein_elastic_2020,pillwein_generalized_2021}.
Bricklaying is an ancient yet still widely used construction approach that uses prefabricated, uniformly sized materials. It can form not only flat surfaces but also curved surfaces \cite{panozzo_designing_2013} and rich textures. This process is also the topic of extensive research in digital construction.

\textbf{Industrial Products.}
Developable surfaces are also widely used in the design and manufacturing of industrial products. Industrial products are functional goods produced in large quantities, and they are standardized. Although industrial products can be manufactured by other processes, production based on developable surfaces can reduce the difficulty of manufacturing and influence the design style of industrial products \cite{chen_fabricable_2017,schuller_shape_2018}.
Vehicle here refers to its shape design and manufacturing. As the shapes of vehicles are often complex surfaces, the direct production of complex surfaces is a costly and complex process. The generation of shapes for automobiles and ships uses the method of combining many scalable surfaces, which can reduce the manufacturing difficulty and cost \cite{liu_industrial_2011,bo_multi-strip_2019}. A new design style is developed through the developable surface method, such as the concept car GINA released by BMW, which has a fabric tensioned structure that can realize easier repairs and take a new aesthetic effect.
In addition, developable surfaces are also used for inflatable structures, such as inflatable dolls \cite{skouras_designing_2014}. In addition, flexible products can be fabricated by fitting complex shapes with developable surfaces, such as fabric toys \cite{schuller_shape_2018} and wearable devices \cite{markvicka_electrodermis_2019} (see Fig. \ref{application}b).
Chairs are often designed in curved shapes to meet the need for people to sit comfortably. Some studies use developable surfaces to construct curved surfaces that conform to sitting posture and form graceful shapes \cite{papanikolaou_bodypods_2015,tang_interactive_2016,vergauwen_computational_2017,kilian_string_2017}.

\textbf{Arts and Crafts.}
Arts and crafts is an artistic activity that creates physical forms with aesthetics through some processes and materials. Compared with painting as fine art, arts and crafts are limited by materials and crafts. The physical limitation also gives arts and crafts special artistic styles. Shapes, generated by skillfully bending and folding, have polygonal and regular features \cite{mundilova_mathematical_2019, jain_hydrogel-based_2021, signer_developable_2021, hafner_design_2021}. Some of the research is inspired by origami, where a shape is fitted by folding a flat surface \cite{tachi_origamizing_2010, akleman_construction_2016, xi_learning_2016, guseinov_curveups_2017, kim_disjoint_2017, jiang_freeform_2020, zhao_developability-driven_2022}(see Fig. \ref{tree}b, \ref{tuck}, \ref{application}c). In addition, some types of forms and modeling methods are enhanced by inspiration of developable surfaces method, such as weaving art with free-form surfaces \cite{ren_3d_2021}, sky lanterns with free-form surfaces \cite{lin_making_2014}, sculpture with fabric mold \cite{zhang_computational_2019}, and peeling art by isometric mapping \cite{liu_computational_2019} (see Fig. \ref{application}d), These art forms based on developable surfaces demonstrate creativity through the rational use of materials.

\textbf{Garments.}
Garments typically consist of many flat fabrics, and this manufacturing process directly affects design process of garment design. Many 3D cutting techniques are put forward for the employment of the bending and folding of developable surfaces in garment design \cite{pietroni_computational_2022}, Designers present the design language by fully applying the characteristics of 3D cutting techniques to achieve artistic shapes. In addition, some studies on 3D fabric manufacturing are based on the method of developable surfaces, which allows the fabric to be produced directly in the 3D form, and sewing is unnecessary \cite{liu_knitting_2021} (see Fig. \ref{fabrication}e).

\textbf{Mechanical Materials.}
In the research for mechanical materials, the bending and folding properties of developable surfaces are used in the study of deformable structures, such as 4D printing materials \cite{tahouni_self-shaping_2020} and design of deformable structural materials \cite{konakovic_beyond_2016, konakovic-lukovic_rapid_2018}. The concept of developable surfaces is also used in the production of folded structural materials by 3D printing \cite{noma_pop-up_2020}. In addition, deformable structure combined with some electro-mechanical control system can realize tangible interaction, resulting in more interactive functions \cite{lee_origami_2018, purnendu_soft_2021, eguchi_pneumatic_2022, yan_fibercuit_2022, yang_compumat_2023, chang_kirigami_2019, kim_expandial_2019, zheng_sensing_2019, chang_kirigami_2020} (see Fig. \ref{application}e). The study of developable surfaces provides novel approaches for the design of new materials and interactions.

\textbf{Data Physicalization.}
Developable surfaces as a type of physical structure provide effective methods for data physicalization. Some studies explored anatomical physical visualization \cite{schindler_anatomical_2020} and design of data physicalization \cite{daneshzand_kiriphys_2022}. The use of developable surfaces has more advantages than static physical objects, such as simple fabrication and dynamic presentation of data.

In conclusion, a wide range of applications is based on developable surfaces, both as a low-cost manufacturing process to optimize production processes and as a way to develop new design processes and design styles. From the above application cases, an innovative idea needs to be paid attention to: the characteristics of developable surfaces are fully considered and appropriate fabrication are used, which often results in innovative applications.

\section{Discussion}\label{sec7}

In previous sections, we summarize the processes of developable surfaces according to our proposed three sections, which are digital modeling, physical modeling, and interactive assistance. We also categorize typical applications based on developable surfaces. In this section, we discuss the challenges retained in the existing research and propose directions for future research.

\begin{figure*}[t!]
    \centering
    \includegraphics[width=\textwidth]{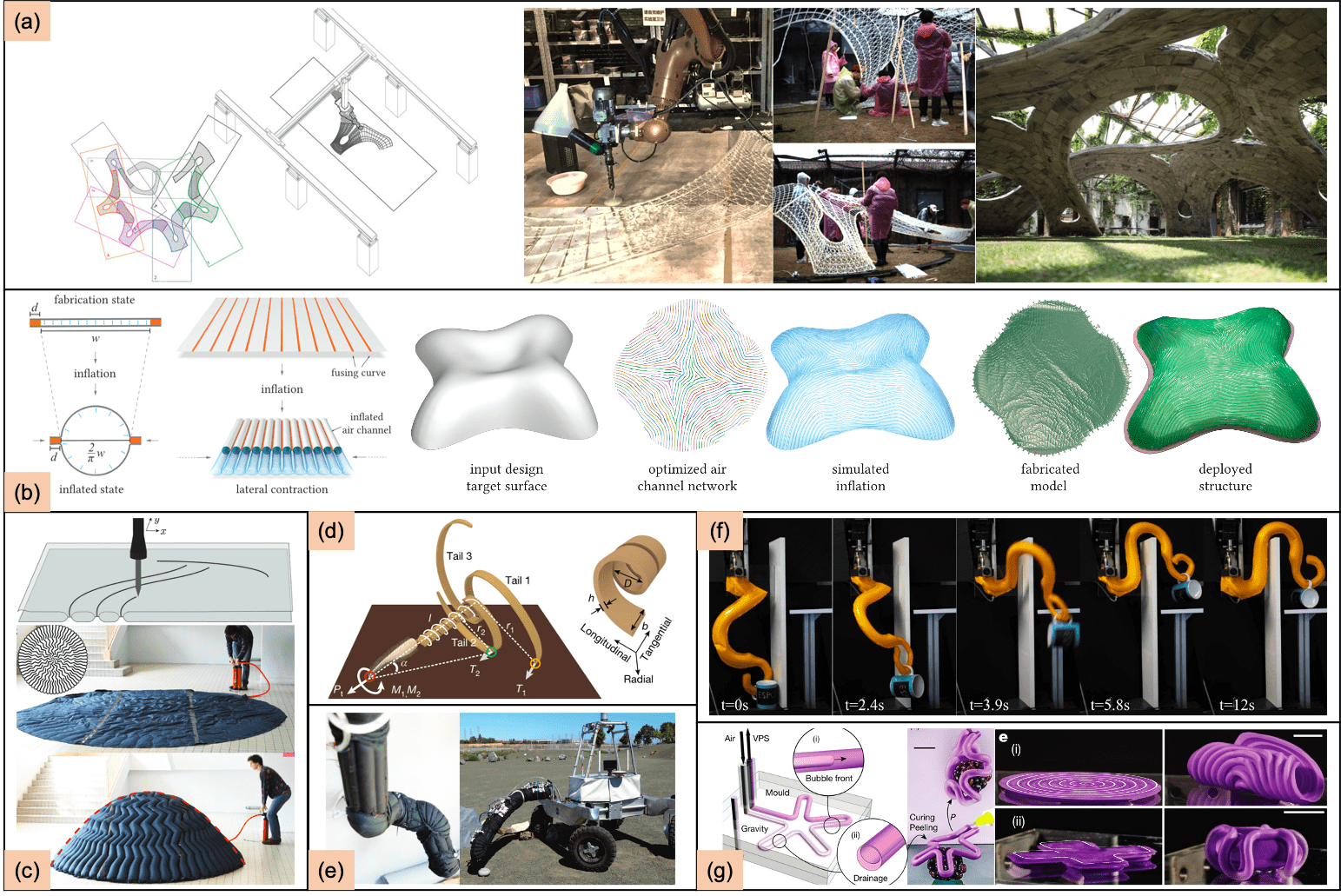}
    \vspace{-20pt}
    \caption{Expanded research based on developable surfaces:
    (a) path printing by robotics for quickly construction of shell \cite{yuan_3d-printed_2020}, (b) a construction method of free-form surface by inflatable and developable structrues \cite{panetta_computational_2021}, (c) inflatable and developable structrues used for regular buildings \cite{siefert_programming_2020}, (d) self-bending for seeding \cite{luo2023autonomous}, (e) inflatable structrues used for soft robotics \cite{bodily_multi-objective_2017}, (f) inflatable and developable structrues used for soft robotics motion \cite{siefert_programming_2019}, (g) a method of fabrication of soft robotics \cite{jones_bubble_2021}.}
    \label{future}
\end{figure*}

\textbf{Quality of Segmentation.}
The quality of segmentation affects subsequent surface flattening, manufacturing, and assembly. Segmentation can be evaluated in several ways, such as whether the parts are regular, similar, and can be easily flattened. Despite the automated segmentation methods available in our collected studies, complicated shapes still require prior guidance from the user. Automated segmentation methods can generate usable results, but often generate irregular results \cite{binninger_developable_2021, zhao_developability-driven_2022}, which increases the difficulty of subsequent manufacturing and assembly. Manual segmentation can generate better results based on experience when dealing with regular shapes, but handling the complicated shapes manually remains a challenge \cite{schuller_shape_2018}. In addition, some general segmentation methods can be explored for complicated shapes according to the collection and analysis of geometric attributes. On the other hand, some new fabrication methods can be developed to make segmentation easier.

\textbf{Complexity of Operation.}
When dealing with special complicated shapes, more accurate cutting solutions and unfolding results can be obtained by interactive operations. The interaction and feedback with users can further optimize the algorithm and improve its reliability and stability. However, the current operation of shape segmentation and unfolding is complicated, and the operator can only follow the rules with restrictions \cite{tang_interactive_2016, vergauwen_computational_2017, schuller_shape_2018, malomo_flexmaps_2018, pietroni_computational_2022}. A possible research can be developed by assistance of computer vision recognizing the state of shapes to simplify manual operations.

\textbf{Data Intelligence.}
Current task of flattening shapes still needs manual aid when dealing with complicated shapes, which increases the difficulty of operation for the subsequent design and manufacture. One approach is similar to semantic-based segmentation \cite{guo_3d_2016, maron_convolutional_2017, herouane_labelling_2018, george_deep_2022} and geometry-based mesh segmentation \cite{poulenard_multi-directional_2018, he_curvanet_2020, milano_primal-dual_2020, lahav_meshwalker_2020, wen_dual-sampling_2021, hu_subdivision-based_2022, sharp_diffusionnet_2022}. Classification models are trained by deep learning for the problem of shape segmentation, and the surface formed by the segmentation is suitable for being unfolded. Another approach is to train efficient surface unfolding models by building large-scale datasets for surface flattening, which provides automated and intelligent solutions to surface unfolding problems.

\textbf{Digitization to Physicalization.}
Most of the current fabrication of developable surfaces is still relies heavily on non-digital experience, and constraints of real materials exist \cite{jiang_freeform_2020, jain_hydrogel-based_2021, signer_developable_2021}, which results in the difference between the digital model and physical model. Many current studies use finite element analysis to simulate material forces to reduce errors before producing physical shapes \cite{peraza_hernandez_smi_2013, erdine_curved-crease_2019, pillwein_elastic_2020, pillwein_generalized_2021}, but this method requires complex calculations and takes a substantial amount of time. A promising approach to addressing the challenge is to train the AI model with data of physical shapes to quickly predict physical-based simulated results \cite{yang_simulearn_2020}. It can bridge the gap between design and manufacturing by enabling more accurate and efficient design-to-manufacture workflows.

\textbf{Future Work.}
The challenges discussed above primarily revolve around technical and operational aspects. However, a more crucial question is the extent to which developable surfaces can solve specific problems and whether they offer greater advantages than other techniques. Thus, it is worth exploring new avenues for studying developable surfaces.

Recent years, the research about "design to fabrication" is getting more and more attention. Owing to conventional workflow from design to fabrication is disjointed, which though is convenient to non-professional users. However, rules and patterns of fabrication (such as G-code) are used in design process, which can solve problems that cannot achieved by conventional fabrication and realize innovations. For examples, quickly construction of large-scale shell by using developable surfaces and planar path printing based on robotics \cite{yuan_3d-printed_2020} (see fig. \ref{future}a). This study provides a idea of combination of developable approximation and planar path printing. In addition, some research explored inflatable and developable structures and the airway paths how to effect deformation from a plane into a surface, which are used for quickly construction of lightweight curved buildings \cite{siefert_programming_2020, panetta_computational_2021} (see fig. \ref{future}b, c). Furthermore some studies have successfully leveraged the benefits of developable surfaces and combined them with other techniques to expand research possibilities. For example, researchers have utilized self-bending wood to facilitate seed burial in the ground\cite{luo2023autonomous}, developable surfaces are transformed into inflatable structures for soft robots\cite{bodily_multi-objective_2017, jones_bubble_2021, siefert_programming_2019}, and inflatable and developable structures are employed for collision protection of drones\cite{nguyen_soft-bodied_2023}.

Some advanced research with combination of material attributes and fabrication is also getting attention. how to design the structures (such as combination of path generation and optimization) to get better material attributes for meeting specific needs (such as self-deformation and stiffness increase), which will offer many potential research opportunities and application value. By employing relevant geometric design, structural optimization, and targeted manufacturing processes, practical problems can be effectively addressed.

\section{Conclusion}\label{sec8}

We summarized the segmentation, optimization, main manufacturing, and assembly methods involved in developable surfaces research based on the collected studies, as well as show the corresponding interactive assistance and design applications. In addition, we summarized a pipeline for the process from design to manufacture of developable surfaces and a visual analysis diagram of the collected studies. Finally, we discussed current challenges in the study of developable surfaces around technical and operational aspects, and suggested opportunities and potential research directions by expanded research.

\section*{ACKNOWLEDGMENTS}
\noindent Nan Cao and Yang Shi are the corresponding authors. This work was supported by NSFC 62072338, 62061136003 and NSF Shanghai 20ZR1461500.

\bibliographystyle{IEEEtran}
\bibliography{reference}

\end{document}